\documentclass[prd,aps,floatfix,superscriptaddress,twocolumn]{revtex4-2}
\usepackage{amsfonts,amsmath,amssymb,amsthm}
\usepackage{bm}
\usepackage{booktabs}
\usepackage{braket} 
\usepackage{cases}
\usepackage{color}
\usepackage{dcolumn}
\usepackage{epsfig}
\usepackage{epstopdf}   
\usepackage{graphicx}
\usepackage[colorlinks,citecolor=blue,anchorcolor=red,menucolor=red,linkcolor=red,filecolor=red,runcolor=red,urlcolor=blue,frenchlinks=red]{hyperref}
\usepackage{indentfirst}
\usepackage{latexsym}
\usepackage{longtable}
\usepackage{multirow}
\usepackage{rotating}
\usepackage{slashed}  
\usepackage{subfigure}
\newcommand{\etal}{\textit{et~al}.~}

\allowdisplaybreaks[4]
\begin{document}
\title{Triply heavy tetraquark states}
\author{Xin-Zhen Weng}
\email{xzhweng@pku.edu.cn}
\affiliation{Center of High Energy Physics, Peking University, Beijing 100871, China}
\affiliation{School of Physics and Astronomy, Tel Aviv University, Tel Aviv 69978, Israel}
\author{Wei-Zhen Deng}
\email{dwz@pku.edu.cn}
\affiliation{School of Physics, Peking University, Beijing 100871, China}
\author{Shi-Lin Zhu}
\email{zhusl@pku.edu.cn}
\affiliation{Center of High Energy Physics, Peking University, Beijing 100871, China}
\date{\today}
\begin{abstract}

In the framework of an extended chromomagnetic model, we
systematically study the mass spectrum of the $S$-wave
$qQ\bar{Q}\bar{Q}$ tetraquarks.
Their mass spectra are mainly determined by the color interaction.
For the $qc\bar{c}\bar{c}$, $qb\bar{c}\bar{c}$ and
$qb\bar{b}\bar{b}$ tetraquarks, the color interaction favors the
color-sextet $\ket{(qQ)^{6_{c}}(\bar{Q}\bar{Q})^{\bar{6}_{c}}}$
configuration over the color-triplet
$\ket{(qQ)^{\bar{3}_{c}}(\bar{Q}\bar{Q})^{3_{c}}}$ one.
But for the $qc\bar{b}\bar{b}$ tetraquarks, the color-triplet
configuration is favored.
We find no stable states which lie below the thresholds of two
pseudoscalar mesons.
The lowest axial-vector states with the $qQ\bar{b}\bar{b}$ flavor
configuration may be narrow. They lie just above the thresholds of
two pseudoscalar mesons, but cannot decay into these channels
because of the conservation of the angular momentum and parity.
%

\end{abstract}

\maketitle
\thispagestyle{empty} 

\section{Introduction}
\label{Sec:Introduction}

Searching for the exotic states is an important and challenging
topic in hadron physics.
In the past two decades, we have witnessed tremendous progresses in
this area.
In 2003, the first charmoniumlike state $X(3872)$ was observed by
the Belle Collaboration~\cite{Choi:2003ue}, and later confirmed by
the CDF~\cite{Acosta:2003zx}, D0~\cite{Abazov:2004kp},
$BABAR$~\cite{Aubert:2004ns}, LHCb~\cite{Aaij:2011sn},
CMS~\cite{Chatrchyan:2013cld}, and BESIII~\cite{Ablikim:2013dyn}
Collaborations.
Its quantum number is $I^{G}J^{PC}=0^{+}1^{++}$~\cite{Zyla:2020zbs}.
Since then, lots of charmoniumlike and bottomoniumlike states are
found.
They are called $XYZ$ states.
Many of them do not fit into the conventional charmonium or
bottomonium spectrum in the quark model.
In particular, the charged charmoniumlike and bottomoniumlike states
like $Z_{c}(3900)$~\cite{Ablikim:2013mio,Liu:2013dau},
$Z_{c}(3885)$~\cite{BESIII:2013qmu,BESIII:2015pqw},
$Z_{c}(4020)$~\cite{BESIII:2013ouc},
$Z_{c}(4025)$~\cite{BESIII:2013mhi},
$Z_{cs}(3985)$~\cite{Ablikim:2020hsk}, $Z_{cs}(4000)$ and
$Z_{cs}$(4020)~\cite{LHCb:2021uow}, $Z_{b}(10610)$ and
$Z_{b}(10650)$~\cite{Bondar:2011aa} contain at least four quarks.
They are candidates of exotic structures like the compact
tetraquark~\cite{Maiani:2004vq,Cui:2006mp,Ebert:2007rn,Park:2013fda,Anwar:2018sol}, the
molecule~\cite{Tornqvist:1993ng,Tornqvist:2004qy,Swanson:2003tb,Hanhart:2007yq,Carames:2010zz,Aceti:2012cb,Chen:2015add}, the
hybrid meson~\cite{Zhu:2005hp,Esposito:2016itg}, and so on.
Interested readers may refer to
Refs.~\cite{Chen:2016qju,Esposito:2016noz,Lebed:2016hpi,Ali:2017jda,Karliner:2017qhf,Olsen:2017bmm,Guo:2017jvc,Yuan:2018inv,Liu:2019zoy,Brambilla:2019esw}
for more details.

Besides the charmoniumlike and bottomoniumlike states, various open
heavy flavor exotic states have been observed recently.
In 2016, the D0 Collaboration reported the observation of the
$X(5568)$ state in the $B_{s}^{0}\pi^{\pm}$
channel~\cite{D0:2016mwd}.
It was very likely a $us\bar{d}\bar{b}$ or $ds\bar{u}\bar{b}$
tetraquarks since it lies about 200~MeV below the $\bar{B}K$
threshold~\cite{Liu:2016ogz}.
However, the subsequent search of the LHCb Collaboration did not
confirm this state~\cite{Aaij:2016iev}.
In 2020, the LHCb Collaboration observed two resonances in the
$D^{-}K^{+}$ channel with spin-$0$ and
spin-$1$~\cite{Aaij:2020hon,Aaij:2020ypa}, whose minimal quark
contents are $ud\bar{s}\bar{c}$.
%
%
%
%
Recently, the LHCb Collaboration observed an narrow state in the
$D^{0}D^{0}\pi^{+}$ mass spectrum just below the $D^{*+}D^{0}$
threshold~\cite{LHCb:2021vvq,LHCb:2021auc}.
%
%
%
This is the first doubly charmed tetraquark $T_{cc}^{+}$ observed in
experiment.
Moreover, in 2020, the LHCb collaboration observed a narrow
structure and a wide structure in the $J/\psi$-pair invariant mass
spectrum in the range of 6.2--7.2~GeV, which could be all-charm
hadrons~\cite{Aaij:2020fnh}.

At this stage, it is natural to expect that the triply charmed
tetraquark(s) may be observed in the very near future.
For the triply charmed system, a molecule configuration requires the
exchange of a $D$ meson.
The corresponding interaction range is $\sim0.1~\mathrm{fm}$, which
is a typical scale of the compact tetraquark.
From another point of view, the $D$-meson exchange interaction is
highly suppressed in the typical scale of the molecule
($\sim\mathrm{fm}$).
In other words, the molecule interpretation is not favored for the
triply heavy system.
If a $qc\bar{c}\bar{c}$ state is observed in experiment, it is very
likely a compact tetraquark state.

However, the theoretical study of this system is relatively
scarce~\cite{SilvestreBrac:1992mv,Silvestre-Brac:1993zem,Cui:2006mp,Chen:2016ont,Jiang:2017tdc,Junnarkar:2018twb,Liu:2019mxw,Xing:2019wil,Hudspith:2020tdf,Lu:2021kut}.
In Ref.~\cite{Chen:2016ont}, Chen~\etal found that some axial-vector
$cc\bar{b}\bar{q}$ states lie below the $B_{c}\bar{D}_{(s)}$
thresholds, which are stable against the strong decay.
A lattice study performed by
Junnarkar~\etal~\cite{Junnarkar:2018twb} gives similar results.
They found that the axial-vector $uc\bar{b}\bar{b}$ and
$sc\bar{b}\bar{b}$ lie below their thresholds by
$6\pm11~\mathrm{MeV}$ and $8\pm3~\mathrm{MeV}$ respectively.
However, a recent study by L\"u~\etal~\cite{Lu:2021kut} predicted
that the corresponding states are about several hundred MeV higher.
More theoretical and experimental studies are needed to have a
better understanding of these systems.

For the triply heavy tetraquarks, the interaction is provided by the
gluon exchange and string confinement.
The resulting interactions include the spin-independent Coulomb-type
interaction, linear confinement and spin-dependent chromomagnetic
interaction, tensor interaction and spin-orbit interaction.
When focusing on the $S$-wave states, the tensor and spin-orbit
interactions can be neglected.
Furthermore, we can simplify the model by integrating out spatial
part of the interaction.
Then the interaction becomes
\begin{equation}
- \sum_{i<j} a_{ij} \bm{F}_{i}\cdot\bm{F}_{j} - \sum_{i<j} v_{ij}
\bm{S}_{i}\cdot\bm{S}_{j} \bm{F}_{i}\cdot\bm{F}_{j}\,.
\end{equation}
This interaction gives good account of the $S$-wave mesons and
baryons~\cite{Weng:2018mmf}.
In this work, we use this interaction to study the triply heavy
tetraquarks.
The paper is organized as follows.
In Sec.~\ref{Sec:Model}, we introduce the extended chromomagnetic
model and construct the wave function bases for the triply heavy
tetraquarks.
Then we discuss the numerical results in Sec.~\ref{Sec:Result}.
We conclude in Sec.~\ref{Sec:Conclusion}.
%

\section{The Extended Chromomagnetic Model}
\label{Sec:Model}


For the $S$-wave tetraquark system, we consider the
chromomagnetic model.
The Hamiltonian
reads~\cite{Chan:1978nk,Fukugita:1978sn,Chao:1979mm,Chao:1980dv,Hogaasen:2013nca,Weng:2018mmf,Weng:2019ynv,Weng:2020jao,Weng:2021hje}
\begin{equation}\label{eqn:ECM}
H
= \sum_{i}m_{i}+H_{\text{CE}}+H_{\text{CM}}
\end{equation}
where $m_i$ is the effective mass of $i$th constituent quark which
consists of the constituent quark mass, the kinetic energy, and so
on, $H_{\text{CE}}$ is the colorelectric (CE) interaction which
includes the color linear confinement and Coulomb-type
interaction~\cite{Hogaasen:2013nca,Weng:2018mmf,Weng:2019ynv,Weng:2020jao,Weng:2021hje}
\begin{equation}\label{eqn:ECM:CE}
H_{\text{CE}}
= - \sum_{i<j} a_{ij} \bm{F}_{i}\cdot\bm{F}_{j}\,,
\end{equation}
and $H_{\text{CM}}$ is the chromomagnetic (CM) interaction from the
one-gluon-exchange
(OGE)~\cite{Jaffe:1976ig,Jaffe:1976ih,DeRujula:1975qlm,Cui:2006mp,Buccella:2006fn,Chen:2016ont,Liu:2019zoy}
\begin{equation}\label{eqn:ECM:CM}
H_{\text{CM}}
= - \sum_{i<j} v_{ij} \bm{S}_{i}\cdot\bm{S}_{j}
\bm{F}_{i}\cdot\bm{F}_{j}\,.
\end{equation}
The coupling constants $a_{ij}$ and
$v_{ij}\propto\braket{\alpha_{s}(r)\delta^{3}(\bm{r})}/m_{i}m_{j}$
depend on the spatial wave function and the constituent quark
masses.
$\bm{S}_{i}=\bm{\sigma}_i/2$ and $\bm{F}_{i}={\bm{\lambda}}_i/2$ are
the quark spin and  color operators.
For the antiquark,
\begin{equation}
\bm{S}_{\bar{q}}=-\bm{S}_{q}^{*}\,, \quad
\bm{F}_{\bar{q}}=-\bm{F}_{q}^{*}\,.
\end{equation}

Since
\begin{align}\label{eqn:m+color=color}
& \sum_{i<j} \left(m_i+m_j\right) \bm{F}_{i}\cdot\bm{F}_{j}
\notag\\
={}& \left(\sum_{i}m_{i}\bm{F}_i\right) \cdot
\left(\sum_{i}\bm{F}_{i}\right) - \frac{4}{3} \sum_{i} m_{i}\,,
\end{align}
and the total color operator $\sum_i\bm{F}_i$ nullifies any
color-singlet physical state, we introduce the quark pair mass
parameter
\begin{equation}\label{eqn:para:color+m}
m_{ij} = \left(m_i+m_j\right) + \frac{4}{3} a_{ij}\,,
\end{equation}
to combine the effective quark mass $m_{i}$ and the color
interaction strength $a_{ij}$.
Then we can rewrite the model Hamiltonian
as~\cite{Weng:2018mmf,Weng:2019ynv,Weng:2020jao,Weng:2021hje}
\begin{equation}\label{eqn:hamiltonian:final}
H= -\frac{3}{4} \sum_{i<j}m_{ij}V^{\text{C}}_{ij} -
\sum_{i<j}v_{ij}V^{\text{CM}}_{ij} \,,
\end{equation}
where
\begin{equation}
V^{\text{C}}_{ij}=\bm{F}_{i}\cdot\bm{F}_{j}\,,
\end{equation}
and
\begin{equation}
V^{\text{CM}}_{ij}=\bm{S}_{i}\cdot\bm{S}_{j}\bm{F}_{i}\cdot\bm{F}_{j}
\end{equation}
are the color and CM interactions between quarks, respectively.
Here, $m_{ij}$ and $v_{ij}$ are unknown parameters.
In Ref.~\cite{Weng:2018mmf}, we have used the conventional mesons
and baryons to fit them.
More precisely, we used the pseudoscalar and vector mesons to extract the parameters $m_{q\bar{q}}$ and $v_{q\bar{q}}$, and used the light and singly heavy baryons to fit the $\{m_{qq},v_{qq}\}$ with at most one heavy quark,  estimated the $m_{QQ}$ and $v_{QQ}$ through a quark model consideration.
Their values are listed in Table~\ref{table:parameter}.
With these parameters, we reproduce the meson and baryon masses with errors mostly within 10~MeV (The only exception is the $\Sigma$ baryon,
whose deviation is 15~MeV.).
We also obtained the $\Xi_{cc}$ baryon mass very close to the LHCb Collaboration's measurement ($M_{\mathrm{th.}}=3633.3\pm9.3~\text{MeV}$ versus $M_{\mathrm{exp.}}=3621.55\pm0.23\pm0.30~\text{MeV}$)~\cite{Aaij:2017ueg,Aaij:2019uaz}.
In Ref.~\cite{Weng:2020jao}, we used these parameters to study the fully heavy tetraquark systems.
We found that for the ground states, the color sextet component is more important than the color triplet one, which is consistent with the dynamical calculations~\cite{Wang:2019rdo,Deng:2020iqw}.
We also studied the hidden-charm pentaquark states, and successfully reproduced the four $P_{c}$ states, $P_{c}(4312)$, $P_{c}(4380)$, $P_{c}(4440)$, and $P_{c}(4450)$~\cite{Aaij:2015tga,Aaij:2019vzc} with these parameters~\cite{Weng:2019ynv}.
In this work, we use the same set of parameters to estimate the
masses of the $S$-wave $qQ\bar{Q}\bar{Q}$ tetraquarks.
%
\begin{table*}
    \centering
    \caption{Parameters of the $q\bar{q}$ and $qq$ pairs~\cite{Weng:2018mmf} (in units of $\text{MeV}$).}
    \label{table:parameter}
    \begin{tabular}{lcccccccccccc}
        \toprule[1pt]
        \toprule[1pt]
        Parameter&$m_{n\bar{n}}$&$m_{n\bar{s}}$&$m_{s\bar{s}}$&$m_{n\bar{c}}$&$m_{s\bar{c}}$&$m_{c\bar{c}}$&$m_{n\bar{b}}$&$m_{s\bar{b}}$&$m_{c\bar{b}}$&$m_{b\bar{b}}$\\
        Value&$615.95$&$794.22$&$936.40$&$1973.22$&$2076.14$&$3068.53$&$5313.35$&$5403.25$&$6322.27$&$9444.97$\\
        Parameter&$v_{n\bar{n}}$&$v_{n\bar{s}}$&$v_{s\bar{s}}$&$v_{n\bar{c}}$&$v_{s\bar{c}}$&$v_{c\bar{c}}$&$v_{n\bar{b}}$&$v_{s\bar{b}}$&$v_{c\bar{b}}$&$v_{b\bar{b}}$\\
        Value&$477.92$&$298.57$&$249.18$&$106.01$&$107.87$&$85.12$&$33.89$&$36.43$&$47.18$&$45.98$\\
        \midrule[1pt]
        Parameter&$m_{nn}$&$m_{ns}$&$m_{ss}$&$m_{nc}$&$m_{sc}$&$m_{cc}$&$m_{nb}$&$m_{sb}$&$m_{cb}$&$m_{b{b}}$\\
        Value&$724.85$&$906.65$&$1049.36$&$2079.96$&$2183.68$&$3171.51$&$5412.25$&$5494.80$&$6416.07$&$9529.57$\\
        Parameter&$v_{n{n}}$&$v_{n{s}}$&$v_{ss}$&$v_{n{c}}$&$v_{s{c}}$&$v_{c{c}}$&$v_{n{b}}$&$v_{s{b}}$&$v_{c{b}}$&$v_{b{b}}$\\
        Value&$305.34$&$212.75$&$195.30$&$62.81$&$70.63$&$56.75$&$19.92$&$8.47$&$31.45$&$30.65$\\
        \bottomrule[1pt]
        \bottomrule[1pt]
    \end{tabular}
\end{table*}
%

\subsection{Wave function}
\label{sec:wavefunc}

Before calculating the tetraquark masses, we need to construct their
wave functions.
In principle, the total wave function is a direct product of the
orbital, color, spin and flavor wave functions.
In this work, we only consider the $S$-wave states, the orbital wave
function is always symmetric.
In the $qq{\otimes}\bar{q}\bar{q}$ configuration, we can construct
the following color-spin wave functions $\{\alpha_{i}^{J}\}$
\begin{enumerate}
    \item $J^{P}=0^{+}$:
    \begin{align}
        &\alpha_{1}^{0}=\ket{\left(q_1q_2\right)_{1}^{6}\left(\bar{q}_{3}\bar{q}_{4}\right)_{1}^{\bar{6}}}_{0},
        \notag\\
        &\alpha_{2}^{0}=\ket{\left(q_1q_2\right)_{0}^{6}\left(\bar{q}_{3}\bar{q}_{4}\right)_{0}^{\bar{6}}}_{0},
        \notag\\
        &\alpha_{3}^{0}=\ket{\left(q_1q_2\right)_{1}^{\bar{3}}\left(\bar{q}_{3}\bar{q}_{4}\right)_{1}^{3}}_{0},
        \notag\\
        &\alpha_{4}^{0}=\ket{\left(q_1q_2\right)_{0}^{\bar{3}}\left(\bar{q}_{3}\bar{q}_{4}\right)_{0}^{3}}_{0},
    \end{align}
    \item $J^{P}=1^{+}$:
    \begin{align}
        &\alpha_{1}^{1}=\ket{\left(q_1q_2\right)_{1}^{6}\left(\bar{q}_{3}\bar{q}_{4}\right)_{1}^{\bar{6}}}_{1},
        \notag\\
        &\alpha_{2}^{1}=\ket{\left(q_1q_2\right)_{1}^{6}\left(\bar{q}_{3}\bar{q}_{4}\right)_{0}^{\bar{6}}}_{1},
        \notag\\
        &\alpha_{3}^{1}=\ket{\left(q_1q_2\right)_{0}^{6}\left(\bar{q}_{3}\bar{q}_{4}\right)_{1}^{\bar{6}}}_{1},
        \notag\\
        &\alpha_{4}^{1}=\ket{\left(q_1q_2\right)_{1}^{\bar{3}}\left(\bar{q}_{3}\bar{q}_{4}\right)_{1}^{3}}_{1},
        \notag\\
        &\alpha_{5}^{1}=\ket{\left(q_1q_2\right)_{1}^{\bar{3}}\left(\bar{q}_{3}\bar{q}_{4}\right)_{0}^{3}}_{1},
        \notag\\
        &\alpha_{6}^{1}=\ket{\left(q_1q_2\right)_{0}^{\bar{3}}\left(\bar{q}_{3}\bar{q}_{4}\right)_{1}^{3}}_{1},
    \end{align}
    \item $J^{P}=2^{+}$:
    \begin{align}
        &\alpha_{1}^{2}=\ket{\left(q_1q_2\right)_{1}^{6}\left(\bar{q}_{3}\bar{q}_{4}\right)_{1}^{\bar{6}}}_{2},
        \notag\\
        &\alpha_{2}^{2}=\ket{\left(q_1q_2\right)_{1}^{\bar{3}}\left(\bar{q}_{3}\bar{q}_{4}\right)_{1}^{3}}_{2},
    \end{align}
\end{enumerate}
where the superscript $3$, $\bar{3}$, $6$ or $\bar{6}$ denotes the
color, and the subscript $0$, $1$ or $2$ denotes the spin.

Next we consider the flavor wave functions.
According to symmetric properties of the two antiquarks, the triply
heavy tetraquarks can be divided into two categories.
The $qQ\bar{c}\bar{b}$ tetraquarks are not constrained by the Pauli
principle, thus all the preceding color-spin wave functions are
allowed.
On the other hand, the $qQ\bar{c}\bar{c}$ and $qQ\bar{b}\bar{b}$
tetraquarks have symmetric flavor wave functions over the two
antiquarks, thus the color-spin wave functions must be asymmetric
over the two antiquarks.
More precisely, the bases for the $qQ\bar{c}\bar{c}$ and
$qQ\bar{b}\bar{b}$ tetraquarks are
$\{\alpha_{2}^{0},\alpha_{3}^{0}\}$,
$\{\alpha_{2}^{1},\alpha_{4}^{1},\alpha_{6}^{1}\}$ and
$\{\alpha_{2}^{2}\}$.

Diagonalizing the Hamiltonian [Eq.~\eqref{eqn:hamiltonian:final}] in
the corresponding bases, we can obtain the masses and eigenvectors
of the triply heavy tetraquarks.
%

\section{Numerical results}
\label{Sec:Result}

\subsection{The $qc\bar{c}\bar{c}$ and $qb\bar{c}\bar{c}$ systems}
\label{Sec:qccc+qbcc}

\begin{figure*}
    \begin{tabular}{ccc}
        \includegraphics[width=450pt]{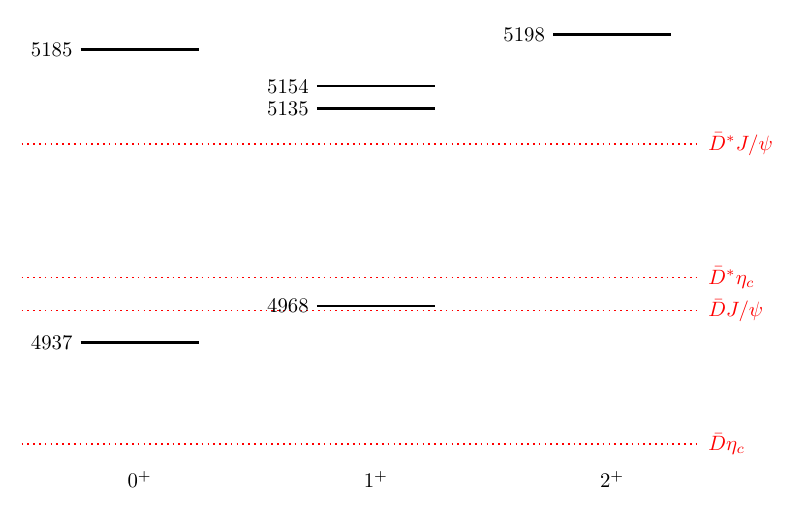}\\
        (a) $nc\bar{c}\bar{c}$ states\\
        &&\\
        \includegraphics[width=450pt]{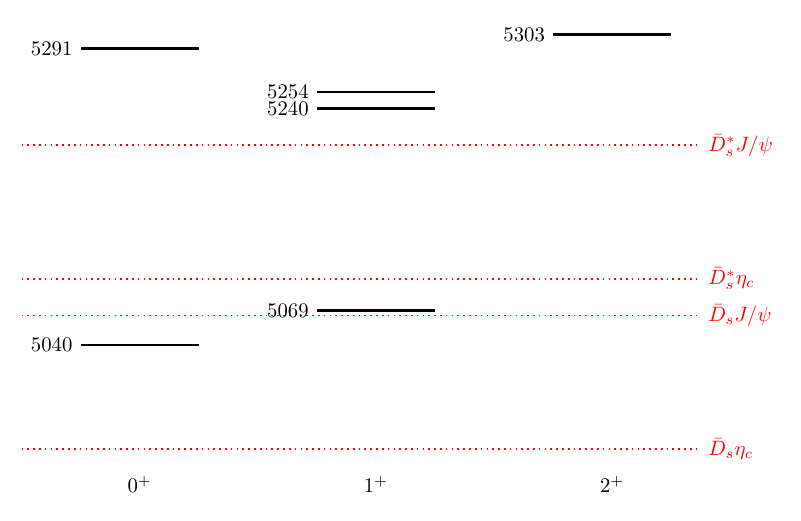}\\
        (b) $sc\bar{c}\bar{c}$ states\\
    \end{tabular}
    \caption{Mass spectra of the $nc\bar{c}\bar{c}$ and $sc\bar{c}\bar{c}$ tetraquark states. The dotted lines indicate various meson-meson thresholds. The masses are all in units of MeV.}
    \label{fig:nccc+sccc}
\end{figure*}
%
\begin{figure*}
    \begin{tabular}{ccc}
        \includegraphics[width=450pt]{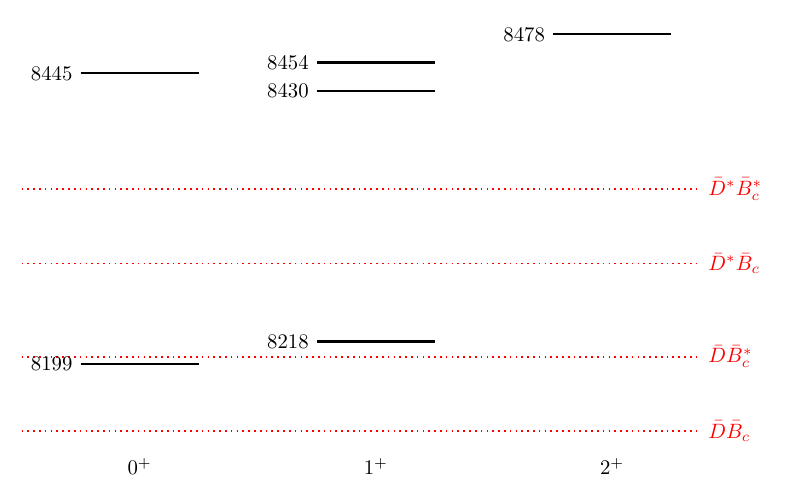}\\
        (a) $nb\bar{c}\bar{c}$ states\\
        &&\\
        \includegraphics[width=450pt]{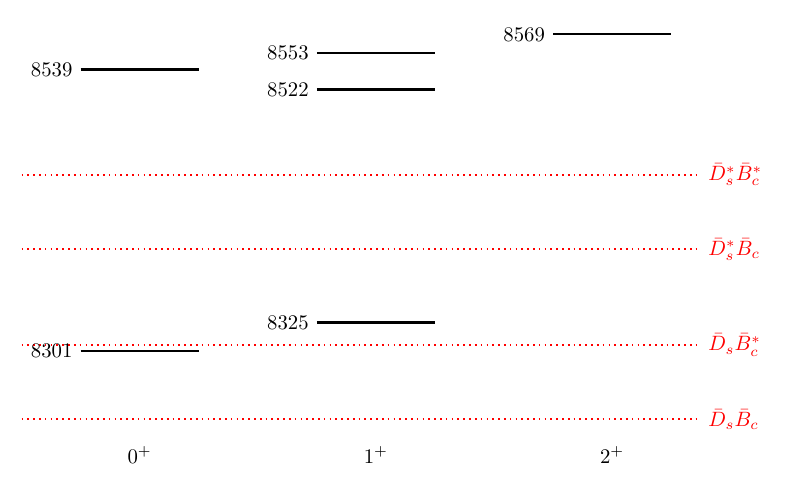}\\
        (b) $sb\bar{c}\bar{c}$ states\\
    \end{tabular}
    \caption{Mass spectra of the $nb\bar{c}\bar{c}$ and $sb\bar{c}\bar{c}$ tetraquark states. The dotted lines indicate various meson-meson thresholds. Here the predicted mass $M_{B_{c}^{*}}=6338~\text{MeV}$ of Godfrey {\it et~al.}~\cite{Godfrey:1985xj} is used. The masses are all in units of MeV.}
    \label{fig:nbcc+sbcc}
\end{figure*}
%

\begin{table}
    \centering
    \caption{Masses and eigenvectors of the $nc\bar{c}\bar{c}$ and $sc\bar{c}\bar{c}$ tetraquarks. The masses are all in units of MeV.}
    \label{table:mass:nccc+sccc}
    \begin{tabular}{ccccccc}
        \toprule[1pt]
        \toprule[1pt]
        System&$J^{P}$&Mass&Eigenvector\\
        \midrule[1pt]
        $nc\bar{c}\bar{c}$&$0^{+}$
        &$4936.7$&$\{0.817,0.576\}$\\
        &
        &$5185.3$&$\{-0.576,0.817\}$\\
        &$1^{+}$
        &$4968.1$&$\{0.915,-0.069,-0.397\}$\\
        &
        &$5135.5$&$\{-0.071,-0.997,0.010\}$\\
        &
        &$5154.0$&$\{-0.397,0.019,-0.918\}$\\
        &$2^{+}$
        &$5198.4$&$\{1\}$\\
        \midrule[1pt]
        $sc\bar{c}\bar{c}$&$0^{+}$
        &$5040.1$&$\{0.816,0.578\}$\\
        &
        &$5290.6$&$\{-0.578,0.816\}$\\
        &$1^{+}$
        &$5069.2$&$\{0.912,-0.074,-0.404\}$\\
        &
        &$5239.9$&$\{-0.076,-0.997,0.011\}$\\
        &
        &$5254.3$&$\{-0.404,0.020,-0.915\}$\\
        &$2^{+}$
        &$5303.3$&$\{1\}$\\
        \bottomrule[1pt]
        \bottomrule[1pt]
    \end{tabular}
\end{table}
%
\begin{table}
    \centering
    \caption{The eigenvectors of the $nc\bar{c}\bar{c}$ tetraquarks in the $n\bar{c}{\otimes}c\bar{c}$ configuration. The masses are all in units of MeV.}
    \label{table:eigenvector:nccc}
    \begin{tabular}{ccccccc}
        \toprule[1pt]
        \toprule[1pt]
        System&$J^{P}$&Mass&$\bar{D}^{*}{J/\psi}$&$\bar{D}^{*}{\eta_{c}}$&$\bar{D}{J/\psi}$&$\bar{D}{\eta_{c}}$\\
        \midrule[1pt]
        $nc\bar{c}\bar{c}$&$0^{+}$
        &$4936.7$&$0.412$&&&$0.622$\\
        &
        &$5185.3$&$-0.643$&&&$0.174$\\
        &$1^{+}$
        &$4968.1$&$0.366$&$0.460$&$-0.516$\\
        &
        &$5135.5$&$-0.037$&$-0.439$&$-0.375$\\
        &
        &$5154.0$&$-0.604$&$0.111$&$-0.095$\\
        &$2^{+}$
        &$5198.4$&$0.577$\\
        \bottomrule[1pt]
        \bottomrule[1pt]
    \end{tabular}
\end{table}
%
\begin{table}
    \centering
    \caption{The eigenvectors of the $sc\bar{c}\bar{c}$ tetraquarks in the $s\bar{c}{\otimes}c\bar{c}$ configuration. The masses are all in units of MeV.}
    \label{table:eigenvector:sccc}
    \begin{tabular}{ccccccc}
        \toprule[1pt]
        \toprule[1pt]
        System&$J^{P}$&Mass&$\bar{D}_{s}^{*}{J/\psi}$&$\bar{D}_{s}^{*}{\eta_{c}}$&$\bar{D}_{s}{J/\psi}$&$\bar{D}_{s}{\eta_{c}}$\\
        \midrule[1pt]
        $sc\bar{c}\bar{c}$&$0^{+}$
        &$5040.1$&$0.410$&&&$0.622$\\
        &
        &$5290.6$&$-0.644$&&&$0.172$\\
        &$1^{+}$
        &$5069.2$&$0.361$&$0.459$&$-0.519$\\
        &
        &$5239.9$&$-0.039$&$-0.441$&$-0.373$\\
        &
        &$5254.3$&$-0.607$&$0.107$&$-0.091$\\
        &$2^{+}$
        &$5303.3$&$0.577$\\
        \bottomrule[1pt]
        \bottomrule[1pt]
    \end{tabular}
\end{table}
%
\begin{table}
    \centering
    \caption{The values of $k\cdot|c_{i}|^2$ for the $nc\bar{c}\bar{c}$ tetraquarks (in unit of MeV).}
    \label{table:kc_i^2:nccc}
    \begin{tabular}{ccccccccccccc}
        \toprule[1pt]
        \toprule[1pt]
        System&$J^{P}$&Mass&$\bar{D}^{*}{J/\psi}$&$\bar{D}^{*}{\eta_{c}}$&$\bar{D}{J/\psi}$&$\bar{D}{\eta_{c}}$\\
        \midrule[1pt]
        $nc\bar{c}\bar{c}$&$0^{+}$
        &$4936.7$&$\times$&&&$172.8$\\
        &
        &$5185.3$&$183.4$&&&$27.0$\\
        &$1^{+}$
        &$4968.1$&$\times$&$\times$&$25.9$\\
        &
        &$5135.5$&$0.4$&$114.1$&$90.0$\\
        &
        &$5154.0$&$125.7$&$7.7$&$6.1$\\
        &$2^{+}$
        &$5198.4$&$159.5$\\
        \bottomrule[1pt]
        \bottomrule[1pt]
    \end{tabular}
\end{table}
%
\begin{table}
    \centering
    \caption{The values of $k\cdot|c_{i}|^2$ for the $sc\bar{c}\bar{c}$ tetraquarks (in unit of MeV).}
    \label{table:kc_i^2:sccc}
    \begin{tabular}{ccccccccccccc}
        \toprule[1pt]
        \toprule[1pt]
        System&$J^{P}$&Mass&$\bar{D}_{s}^{*}{J/\psi}$&$\bar{D}_{s}^{*}{\eta_{c}}$&$\bar{D}_{s}{J/\psi}$&$\bar{D}_{s}{\eta_{c}}$\\
        \midrule[1pt]
        $sc\bar{c}\bar{c}$&$0^{+}$
        &$5040.1$&$\times$&&&$178.2$\\
        &
        &$5290.6$&$188.8$&&&$27.0$\\
        &$1^{+}$
        &$5069.2$&$\times$&$\times$&$26.4$\\
        &
        &$5239.9$&$0.4$&$117.3$&$91.1$\\
        &
        &$5254.3$&$124.4$&$7.3$&$5.6$\\
        &$2^{+}$
        &$5303.3$&$163.1$\\
        \bottomrule[1pt]
        \bottomrule[1pt]
    \end{tabular}
\end{table}
%
\begin{table}
    \centering
    \caption{The partial width ratios for the $nc\bar{c}\bar{c}$ tetraquarks. For each state, we choose one mode as the reference channel, and the partial width ratios of the other channels are calculated relative to this channel. The masses are all in unit of MeV.}
    \label{table:R:nccc}
    \begin{tabular}{ccccccccccccc}
        \toprule[1pt]
        \toprule[1pt]
        System&$J^{P}$&Mass&$\bar{D}^{*}{J/\psi}$&$\bar{D}^{*}{\eta_{c}}$&$\bar{D}{J/\psi}$&$\bar{D}{\eta_{c}}$\\
        \midrule[1pt]
        $nc\bar{c}\bar{c}$&$0^{+}$
        &$4936.7$&$\times$&&&$1$\\
        &
        &$5185.3$&$6.8$&&&$1$\\
        &$1^{+}$
        &$4968.1$&$\times$&$\times$&$1$\\
        &
        &$5135.5$&$0.004$&$1.3$&$1$\\
        &
        &$5154.0$&$20.7$&$1.3$&$1$\\
        &$2^{+}$
        &$5198.4$&$1$\\
        \bottomrule[1pt]
        \bottomrule[1pt]
    \end{tabular}
\end{table}
%
\begin{table}
    \centering
    \caption{The partial width ratios for the $sc\bar{c}\bar{c}$ tetraquarks. For each state, we choose one mode as the reference channel, and the partial width ratios of the other channels are calculated relative to this channel. The masses are all in unit of MeV.}
    \label{table:R:sccc}
    \begin{tabular}{ccccccccccccc}
        \toprule[1pt]
        \toprule[1pt]
        System&$J^{P}$&Mass&$\bar{D}_{s}^{*}{J/\psi}$&$\bar{D}_{s}^{*}{\eta_{c}}$&$\bar{D}_{s}{J/\psi}$&$\bar{D}_{s}{\eta_{c}}$\\
        \midrule[1pt]
        $sc\bar{c}\bar{c}$&$0^{+}$
        &$5040.1$&$\times$&&&$1$\\
        &
        &$5290.6$&$7.0$&&&$1$\\
        &$1^{+}$
        &$5069.2$&$\times$&$\times$&$1$\\
        &
        &$5239.9$&$0.005$&$1.3$&$1$\\
        &
        &$5254.3$&$22.1$&$1.3$&$1$\\
        &$2^{+}$
        &$5303.3$&$1$\\
        \bottomrule[1pt]
        \bottomrule[1pt]
    \end{tabular}
\end{table}
%

\begin{table}
    \centering
    \caption{Masses and eigenvectors of the $nb\bar{c}\bar{c}$ and $sb\bar{c}\bar{c}$ tetraquarks. The masses are all in units of MeV.}
    \label{table:mass:nbcc+sbcc}
    \begin{tabular}{ccccccc}
        \toprule[1pt]
        \toprule[1pt]
        System&$J^{P}$&Mass&Eigenvector\\
        \midrule[1pt]
        $nb\bar{c}\bar{c}$&$0^{+}$
        &$8199.1$&$\{0.907,0.421\}$\\
        &
        &$8444.8$&$\{-0.421,0.907\}$\\
        &$1^{+}$
        &$8217.5$&$\{0.958,-0.151,-0.244\}$\\
        &
        &$8430.0$&$\{0.207,0.953,0.222\}$\\
        &
        &$8454.4$&$\{-0.199,0.263,-0.944\}$\\
        &$2^{+}$
        &$8477.9$&$\{1\}$\\
        \midrule[1pt]
        $sb\bar{c}\bar{c}$&$0^{+}$
        &$8300.5$&$\{0.895,0.445\}$\\
        &
        &$8538.7$&$\{-0.445,0.895\}$\\
        &$1^{+}$
        &$8324.8$&$\{0.952,-0.169,-0.256\}$\\
        &
        &$8521.9$&$\{0.211,0.967,0.145\}$\\
        &
        &$8553.3$&$\{-0.223,0.192,-0.956\}$\\
        &$2^{+}$
        &$8569.1$&$\{1\}$\\
        \bottomrule[1pt]
        \bottomrule[1pt]
    \end{tabular}
\end{table}
%
\begin{table}
    \centering
    \caption{The eigenvectors of the $nb\bar{c}\bar{c}$ tetraquarks in the $n\bar{c}{\otimes}b\bar{c}$ configuration. The masses are all in units of MeV.}
    \label{table:eigenvector:nbcc}
    \begin{tabular}{ccccccc}
        \toprule[1pt]
        \toprule[1pt]
        System&$J^{P}$&Mass&$\bar{D}^{*}\bar{B}_{c}^{*}$&$\bar{D}^{*}\bar{B}_{c}$&$\bar{D}\bar{B}_{c}^{*}$&$\bar{D}\bar{B}_{c}$\\
        \midrule[1pt]
        $nb\bar{c}\bar{c}$&$0^{+}$
        &$8199.1$&$0.520$&&&$0.581$\\
        &
        &$8444.8$&$-0.559$&&&$0.282$\\
        &$1^{+}$
        &$8217.5$&$0.454$&$0.400$&$-0.523$\\
        &
        &$8430.0$&$0.210$&$0.409$&$0.369$\\
        &
        &$8454.4$&$-0.500$&$0.299$&$-0.084$\\
        &$2^{+}$
        &$8477.9$&$0.577$\\
        \bottomrule[1pt]
        \bottomrule[1pt]
    \end{tabular}
\end{table}
%
\begin{table}
    \centering
    \caption{The eigenvectors of the $sb\bar{c}\bar{c}$ tetraquarks in the $s\bar{c}{\otimes}b\bar{c}$ configuration. The masses are all in units of MeV.}
    \label{table:eigenvector:sbcc}
    \begin{tabular}{ccccccc}
        \toprule[1pt]
        \toprule[1pt]
        System&$J^{P}$&Mass&$\bar{D}_{s}^{*}\bar{B}_{c}^{*}$&$\bar{D}_{s}^{*}\bar{B}_{c}$&$\bar{D}_{s}\bar{B}_{c}^{*}$&$\bar{D}_{s}\bar{B}_{c}$\\
        \midrule[1pt]
        $sb\bar{c}\bar{c}$&$0^{+}$
        &$8300.5$&$0.505$&&&$0.588$\\
        &
        &$8538.7$&$-0.573$&&&$0.266$\\
        &$1^{+}$
        &$8324.8$&$0.445$&$0.394$&$-0.532$\\
        &
        &$8521.9$&$0.181$&$0.439$&$0.350$\\
        &
        &$8553.3$&$-0.519$&$0.263$&$-0.106$\\
        &$2^{+}$
        &$8569.1$&$0.577$\\
        \bottomrule[1pt]
        \bottomrule[1pt]
    \end{tabular}
\end{table}
%
\begin{table}
    \centering
    \caption{The values of $k\cdot|c_{i}|^2$ for the $nb\bar{c}\bar{c}$ tetraquarks (in unit of MeV).}
    \label{table:kc_i^2:nbcc}
    \begin{tabular}{ccccccccccccc}
        \toprule[1pt]
        \toprule[1pt]
        System&$J^{P}$&Mass&$\bar{D}^{*}\bar{B}_{c}^{*}$&$\bar{D}^{*}\bar{B}_{c}$&$\bar{D}\bar{B}_{c}^{*}$&$\bar{D}\bar{B}_{c}$\\
        \midrule[1pt]
        $nb\bar{c}\bar{c}$&$0^{+}$
        &$8199.1$&$\times$&&&$137.0$\\
        &
        &$8444.8$&$172.6$&&&$75.9$\\
        &$1^{+}$
        &$8217.5$&$\times$&$\times$&$51.7$\\
        &
        &$8430.0$&$22.4$&$112.9$&$111.6$\\
        &
        &$8454.4$&$144.5$&$65.2$&$6.1$\\
        &$2^{+}$
        &$8477.9$&$213.0$\\
        \bottomrule[1pt]
        \bottomrule[1pt]
    \end{tabular}
\end{table}
%
\begin{table}
    \centering
    \caption{The values of $k\cdot|c_{i}|^2$ for the $sb\bar{c}\bar{c}$ tetraquarks (in unit of MeV).}
    \label{table:kc_i^2:sbcc}
    \begin{tabular}{ccccccccccccc}
        \toprule[1pt]
        \toprule[1pt]
        System&$J^{P}$&Mass&$\bar{D}_{s}^{*}\bar{B}_{c}^{*}$&$\bar{D}_{s}^{*}\bar{B}_{c}$&$\bar{D}_{s}\bar{B}_{c}^{*}$&$\bar{D}_{s}\bar{B}_{c}$\\
        \midrule[1pt]
        $sb\bar{c}\bar{c}$&$0^{+}$
        &$8300.5$&$\times$&&&$143.8$\\
        &
        &$8538.7$&$175.2$&&&$68.0$\\
        &$1^{+}$
        &$8324.8$&$\times$&$\times$&$66.7$\\
        &
        &$8521.9$&$15.7$&$126.8$&$100.4$\\
        &
        &$8553.3$&$155.2$&$50.7$&$9.9$\\
        &$2^{+}$
        &$8569.1$&$206.3$\\
        \bottomrule[1pt]
        \bottomrule[1pt]
    \end{tabular}
\end{table}
%
\begin{table}
    \centering
    \caption{The partial width ratios for the $nb\bar{c}\bar{c}$ tetraquarks. For each state, we choose one mode as the reference channel, and the partial width ratios of the other channels are calculated relative to this channel. The masses are all in unit of MeV.}
    \label{table:R:nbcc}
    \begin{tabular}{ccccccccccccc}
        \toprule[1pt]
        \toprule[1pt]
        System&$J^{P}$&Mass&$\bar{D}^{*}\bar{B}_{c}^{*}$&$\bar{D}^{*}\bar{B}_{c}$&$\bar{D}\bar{B}_{c}^{*}$&$\bar{D}\bar{B}_{c}$\\
        \midrule[1pt]
        $nb\bar{c}\bar{c}$&$0^{+}$
        &$8199.1$&$\times$&&&$1$\\
        &
        &$8444.8$&$2.3$&&&$1$\\
        &$1^{+}$
        &$8217.5$&$\times$&$\times$&$1$\\
        &
        &$8430.0$&$0.2$&$1.01$&$1$\\
        &
        &$8454.4$&$23.8$&$10.7$&$1$\\
        &$2^{+}$
        &$8477.9$&$1$\\
        \bottomrule[1pt]
        \bottomrule[1pt]
    \end{tabular}
\end{table}
%
\begin{table}
    \centering
    \caption{The partial width ratios for the $sb\bar{c}\bar{c}$ tetraquarks. For each state, we choose one mode as the reference channel, and the partial width ratios of the other channels are calculated relative to this channel. The masses are all in unit of MeV.}
    \label{table:R:sbcc}
    \begin{tabular}{ccccccccccccc}
        \toprule[1pt]
        \toprule[1pt]
        System&$J^{P}$&Mass&$\bar{D}_{s}^{*}\bar{B}_{c}^{*}$&$\bar{D}_{s}^{*}\bar{B}_{c}$&$\bar{D}_{s}\bar{B}_{c}^{*}$&$\bar{D}_{s}\bar{B}_{c}$\\
        \midrule[1pt]
        $sb\bar{c}\bar{c}$&$0^{+}$
        &$8300.5$&$\times$&&&$1$\\
        &
        &$8538.7$&$2.6$&&&$1$\\
        &$1^{+}$
        &$8324.8$&$\times$&$\times$&$1$\\
        &
        &$8521.9$&$0.2$&$1.3$&$1$\\
        &
        &$8553.3$&$15.7$&$5.1$&$1$\\
        &$2^{+}$
        &$8569.1$&$1$\\
        \bottomrule[1pt]
        \bottomrule[1pt]
    \end{tabular}
\end{table}

First we consider the $nc\bar{c}\bar{c}$ and $sc\bar{c}\bar{c}$
tetraquarks.
Their masses and eigenvectors are listed in
Tables~\ref{table:mass:nccc+sccc}.
For simplicity, we will use $T(qQ\bar{Q}\bar{Q},m,J^{P})$ to denote
the $qQ\bar{Q}\bar{Q}$ tetraquarks in the following.
The isospin of the tetraquarks can be easily identified with either
$q=n$ or $q=s$.
Here we assume that the $\mathrm{SU}(2)$ flavor symmetry is exact
and $n=\{u,d\}$.
We plot their relative position in Fig~\ref{fig:nccc+sccc}, along
with the corresponding meson-meson thresholds.
From the figure, we can easily see that the quantum number of the
ground states are $J^{P}=0^{+}$ in both cases.
They are $T(nc\bar{c}\bar{c},4936.7,0^{+})$ and
$T(sc\bar{c}\bar{c},5040.1,0^{+})$, respectively.
On the other hand, the highest mass states have quantum number
$J^{P}=2^{+}$.
Comparing to the thresholds, we find that all states are above the
meson-meson thresholds and can decay through $S$-wave.
They may all be \emph{broad} states~\cite{Jaffe:1976ig}.
The lowest axial-vector states
$T(nc\bar{c}\bar{c},4968.1,1^{+})$/$T(sc\bar{c}\bar{c},5069.2,1^{+})$
lie just above the $D_{(s)}J/\psi$ threshold.
Thus they might be relatively narrow compared to other states.

Besides the masses, the eigenvectors also provide important
information of the tetraquarks.
The color configurations of the tetraquarks can be of
$\ket{(q_{1}q_{2})^{6_{c}}(\bar{q}_{3}\bar{q}_{4})^{\bar{6}_{c}}}$
or
$\ket{(q_{1}q_{2})^{\bar{3}_{c}}(\bar{q}_{3}\bar{q}_{4})^{3_{c}}}$.
For simplicity, we denote them as $6_{c}\otimes\bar{6}_{c}$ and
$\bar{3}_{c}\otimes3_{c}$.
From Table~\ref{table:mass:nccc+sccc}, we find that the ground
states of the $nc\bar{c}\bar{c}$ and $sc\bar{c}\bar{c}$ tetraquarks
are both dominated by the color-sextet components.
The $T(nc\bar{c}\bar{c},4936.7,0^{+})$ has $66.7\%$ of the
$6_{c}\otimes\bar{6}_{c}$ component and the
$T(sc\bar{c}\bar{c},5040.1,0^{+})$ has $66.6\%$.
This is very similar to the fully charm
tetraquarks~\cite{Wang:2019rdo,Deng:2020iqw,Lu:2020cns,Weng:2020jao}.
In the one-gluon-exchange (OGE) model, the color interactions are
attractive inside
$(q_{1}q_{2})^{\bar{3}_{c}}$/$(\bar{q}_{3}\bar{q}_{4})^{3_{c}}$ and
repulsive inside
$(q_{1}q_{2})^{6_{c}}$/$(\bar{q}_{3}\bar{q}_{4})^{\bar{6}_{c}}$.
However, the attractions between $(q_{1}q_{2})^{6_{c}}$ and
$(\bar{q}_{3}\bar{q}_{4})^{\bar{6}_{c}}$ is much stronger than that
between $(q_{1}q_{2})^{\bar{3}_{c}}$ and
$(\bar{q}_{3}\bar{q}_{4})^{3_{c}}$.
In our cases, the two competing effects result in the lower mass of
the color-sextet configurations.
More precisely,
%
%
\begin{align}\label{eqn:color:qccc:spin0}
&\Braket{H_{\text{C}}\left(qc\bar{c}\bar{c}\right)}
\notag\\
={}& -\frac{3}{4} \left\langle m_{qc}V_{12}^{\text{C}}
+m_{cc}V_{34}^{\text{C}}
+m_{q\bar{c}}\left(V_{13}^{\text{C}}+V_{14}^{\text{C}}\right)
\right.
\notag\\
&\qquad \left.
+m_{c\bar{c}}\left(V_{23}^{\text{C}}+V_{24}^{\text{C}}\right)
\right\rangle
\notag\\
={}& m_{q\bar{c}} + m_{c\bar{c}} - \frac{3}{2} \delta{m}
\Braket{V_{12}^{\text{C}}+V_{34}^{\text{C}}}
\notag\\
={}& m_{q\bar{c}} + m_{c\bar{c}} + \delta{m}
\begin{pmatrix}
-1&0\\
0&+2
\end{pmatrix}
\end{align}
%
%
where $\delta{m}=(m_{qc}+m_{cc}-m_{q\bar{c}}-m_{c\bar{c}})/4$.
Inserting the parameters in Table~\ref{table:parameter}, we have
\begin{equation}
\delta{m}\left(nc\bar{c}\bar{c}\right)=52.43~\text{MeV}\,,
\end{equation}
\begin{equation}
\delta{m}\left(sc\bar{c}\bar{c}\right)=52.63~\text{MeV}\,.
\end{equation}
Similar to the fully charm cases, the color interaction renders the
$6_{c}\otimes\bar{6}_{c}$ configuration more stable than the
$\bar{3}_{c}\otimes3_{c}$ one.
We also find that the color interaction does not mix the two color
configurations.
This is because the two antiquarks are
identical~\cite{Wang:2019rdo,Weng:2021hje}.
However, the two color configuration will mix due to the CM
interaction
\begin{align}
&\Braket{H_{\text{CM}}\left(qc\bar{c}\bar{c}\right)}
\notag\\
={}&
\begin{pmatrix}
\frac{1}{4}\left(v_{qc}+v_{cc}\right)&-\frac{\sqrt{6}}{4}\left(v_{q\bar{c}}+v_{c\bar{c}}\right)\\
-\frac{\sqrt{6}}{4}\left(v_{q\bar{c}}+v_{c\bar{c}}\right)&\frac{1}{6}\left(v_{qc}+v_{cc}\right)-\frac{1}{3}\left(v_{q\bar{c}}+v_{c\bar{c}}\right)\\
\end{pmatrix}
\notag\\
\approx{}& \frac{v_{q\bar{c}}+v_{c\bar{c}}}{2}
\begin{pmatrix}
\frac{1}{3}&-\sqrt{\frac{3}{2}}\\
-\sqrt{\frac{3}{2}}&-\frac{4}{9}
\end{pmatrix}
\end{align}
where we have used $v_{q_{1}q_{2}}/v_{q_{1}\bar{q}_{2}}\approx2/3$
in the last line~\cite{Weng:2018mmf}.
Here, $(v_{n\bar{c}}+v_{c\bar{c}})/2=95.57~\text{MeV}$ and
$(v_{s\bar{c}}+v_{c\bar{c}})/2=96.50~\text{MeV}$.
The diagonal part of the CM interaction favors the
$\bar{3}_{c}\otimes3_{c}$ configuration by approximately
$75~\text{MeV}$.
However, Eq.\eqref{eqn:color:qccc:spin0} indicates that the color
interaction disfavor the $\bar{3}_{c}\otimes3_{c}$ configuration by
approximately $150~\text{MeV}$.
Thus the ground states of the scalar $qc\bar{c}\bar{c}$ tetraquarks
are dominated by the $6_{c}\otimes\bar{6}_{c}$ component, while the
higher scalar states mostly consist of $\bar{3}_{c}\otimes3_{c}$
component.

There are three bases for the axial-vector states, namely
$\ket{(qc)^{6}_{1}(\bar{c}\bar{c})^{\bar{6}}_{0}}$,
$\ket{(qc)^{\bar{3}}_{1}(\bar{c}\bar{c})^{3}_{1}}$ and
$\ket{(qc)^{\bar{3}}_{0}(\bar{c}\bar{c})^{3}_{1}}$.
In these bases, we have
\begin{equation}\label{eqn:color:qccc:spin2}
\Braket{H_{\text{C}}\left(qc\bar{c}\bar{c}\right)}
= m_{q\bar{c}} + m_{c\bar{c}} + \delta{m}
\begin{pmatrix}
-1&0&0\\
0&+2&0\\
0&0&+2
\end{pmatrix}
\end{equation}
The color interaction splits the three bases into two energy bands.
The $6_{c}\otimes\bar{6}_{c}$ configuration is more stable than the
$\bar{3}_{c}\otimes3_{c}$.
The CM interaction further splits the two bands into the three bands
structure in Fig.~\ref{fig:nccc+sccc}.
As a consequence, the lightest axial-vector states of both
$nc\bar{c}\bar{c}$ and $sc\bar{c}\bar{c}$ tetraquarks have more than
$80\%$ of the $6_{c}\otimes\bar{6}_{c}$ component.
Combining the scalar and axial-vector cases, we can conclude that
the tetraquark spectrum is dominantly determined by the color
interaction, and the CM interaction contributes to the finer
structures.

The eigenvectors can also be used to study the decay properties of
the tetraquarks.
We can transform the wave functions into the
$q\bar{q}{\otimes}q\bar{q}$ configuration.
The corresponding color configuration can be either of
$\ket{(q\bar{q})^{1_{c}}(q\bar{q})^{1_{c}}}$ or
$\ket{(q\bar{q})^{8_{c}}(q\bar{q})^{8_{c}}}$.
The former one can easily decay into two $S$-wave mesons in $S$ wave
(the so-called ``Okubo-Zweig-Iizuka- (OZI-)superallowed'' decays),
while the latter one can fall apart only through gluon exchange.
Following Refs.~\cite{Jaffe:1976ig,Jaffe:1976ih,Strottman:1979qu},
we only consider the ``OZI-superallowed'' decays.
In
Tables~\ref{table:eigenvector:nccc}--\ref{table:eigenvector:sccc},
we transform the eigenvectors of the $qc\bar{c}\bar{c}$ tetraquarks
into the $q\bar{c}{\otimes}c\bar{c}$ configuration.
For simplicity, we only present the color-singlet components, and we
rewrite the bases as a direct product of two mesons.
For each decay channel, the decay width is proportional to the
square of the coefficient $c_{i}$ of the corresponding component in
the eigenvectors, and also depends on the phase space.
For two body decay through $L$-wave, the partial decay width
reads~\cite{Gao-1992-Group,Weng:2019ynv}
\begin{equation}\label{eqn:width}
\Gamma_{i}=\gamma_{i}\alpha\frac{k^{2L+1}}{m^{2L}}{\cdot}|c_i|^2,
\end{equation}
where $\gamma_{i}$ is a quantity determined by the decay dynamics,
$\alpha$ is an effective coupling constant, $k$ is the momentum of
the final states in the rest frame of the initial state, and $m$ is
the mass of the initial state.
In this work, the $(k/m)^2$ is always of $\mathcal{O}(10^{-2})$ or
even smaller.
Thus we will only consider the $S$-wave decays.
%
Next we consider the $\gamma_{i}$.
Generally, $\gamma_{i}$ depends on the spatial wave functions of the
initial tetraquark and final mesons, which are different for each
decay process.
In the quark model, the spatial wave functions of the ground state
scalar and axial-vector meson are the
same~\cite{Eichten:1978tg,Eichten:1979ms}.
Thus for each tetraquark, we have
\begin{equation}
\gamma_{M_{1}M_{2}} = \gamma_{M_{1}M_{2}^{*}} =
\gamma_{M_{1}^{*}M_{2}} = \gamma_{M_{1}^{*}M_{2}^{*}}
\end{equation}
where $M_{i}$ and $M_{i}^{*}$ denote the pseudoscalar and
axial-vector mesons.
With the eigenvectors obtained, we calculate the value of
$k\cdot|c_i|^2$ and the relative widths for the $qc\bar{c}\bar{c}$
decays, as shown in
Tables~\ref{table:kc_i^2:nccc}--\ref{table:R:sccc}.
Their ground states can easily decay into $\bar{D}_{(s)}J/\psi$,
thus may be \emph{broad} states~\cite{Jaffe:1976ig}.
The higher scalar states,
$T(nc\bar{c}\bar{c},5185.3,0^{+})$/$T(sc\bar{c}\bar{c},5290.6,0^{+})$,
can decay to $\bar{D}_{(s)}^{*}J/\psi$ and $\bar{D}_{(s)}J/\psi$
modes through $S$-wave, with relative decay width ratios
\begin{equation}
\frac{\Gamma\left[T(nc\bar{c}\bar{c},5185.3,0^{+}){\to}\bar{D}^{*}J/\psi\right]}{\Gamma\left[T(nc\bar{c}\bar{c},5185.3,0^{+}){\to}\bar{D}J/\psi\right]}
\sim 6.8\,,
\end{equation}
and
\begin{equation}
\frac{\Gamma\left[T(sc\bar{c}\bar{c},5290.6,0^{+}){\to}\bar{D}_{s}^{*}J/\psi\right]}{\Gamma\left[T(sc\bar{c}\bar{c},5290.6,0^{+}){\to}\bar{D}_{s}J/\psi\right]}
\sim 7.0\,.
\end{equation}
The dominant decay modes are the $\bar{D}_{(s)}^{*}J/\psi$ final
states.
Next we consider the axial-vector states.
The $T(nc\bar{c}\bar{c},5135.3,1^{+})$ decay dominantly to
$\bar{D}^{*}\eta_{c}$ and $\bar{D}J/\psi$
\begin{equation}
\Gamma_{\bar{D}^{*}J/\psi}:\Gamma_{\bar{D}^{*}\eta_{c}}:\Gamma_{\bar{D}J/\psi}
\sim 0.004:1.3:1\,.
\end{equation}
And for $T(nc\bar{c}\bar{c},5154.0,1^{+})$
\begin{equation}
\Gamma_{\bar{D}^{*}J/\psi}:\Gamma_{\bar{D}^{*}\eta_{c}}:\Gamma_{\bar{D}J/\psi}
\sim 20.7:1.3:1\,.
\end{equation}
Its dominant decay mode is $\bar{D}^{*}J/\psi$.
The decay property of the $sc\bar{c}\bar{c}$ tetraquarks is similar
the $nc\bar{c}\bar{c}$ tetraquarks.
%


Replacing the charm quark by a bottom quark, we get the
$nb\bar{c}\bar{c}$ and $sb\bar{c}\bar{c}$ tetraquarks.
We list their masses and eigenvectors in
Table~\ref{table:mass:nbcc+sbcc} and plot their relative position in
Fig.~\ref{fig:nbcc+sbcc}, along with the possible decay channels.
Since the $B_{c}^{*}$ meson has not been observed, we use the
Godfrey-Isgur (GI) model's prediction
$M_{B_{c}^{*}}=6338~\text{MeV}$~\cite{Godfrey:1985xj} to estimate
the meson-meson thresholds.
The $qb\bar{c}\bar{c}$ tetraquarks are very similar to the
$qc\bar{c}\bar{c}$ tetraquarks.
The ground states are both scalar.
They are $T(nb\bar{c}\bar{c},8199.1,0^{+})$ and
$T(sb\bar{c}\bar{c},8300.5,0^{+})$, whose color configurations are
$6_{c}\otimes\bar{6}_{c}$ as well ($>80\%$).
The lightest axial-vector states, $T(nb\bar{c}\bar{c},8217.5,1^{+})$
and $T(sb\bar{c}\bar{c},8324.8,1^{+})$ are also dominated by
$6_{c}\otimes\bar{6}_{c}$ component ($>90\%$).
All higher states mainly consist of $\bar{3}_{c}\otimes3_{c}$
component.

The decay properties of the $qb\bar{c}\bar{c}$ tetraquarks can be
found from Tables~\ref{table:eigenvector:nbcc}--\ref{table:R:sbcc}.
Similar to the $qc\bar{c}\bar{c}$ tetraquarks, the two lowest states
are just above their $S$-wave decay channels, while all other states
are above all meson-meson thresholds.
The higher scalar states can decay into
$\bar{D}_{(s)}^{*}\bar{B}_{c}^{*}$ and $\bar{D}_{(s)}\bar{B}_{c}$
channels with  into comparable partial widths
\begin{equation}
\frac{\Gamma\left[T(nb\bar{c}\bar{c},8444.8,0^{+}){\to}\bar{D}^{*}\bar{B}_{c}^{*}\right]}{\Gamma\left[T(nb\bar{c}\bar{c},8444.8,0^{+}){\to}\bar{D}\bar{B}_{c}\right]}
\sim 2.3\,,
\end{equation}
and
\begin{equation}
\frac{\Gamma\left[T(sb\bar{c}\bar{c},8538.7,0^{+}){\to}\bar{D}_{s}^{*}\bar{B}_{c}^{*}\right]}{\Gamma\left[T(sb\bar{c}\bar{c},8538.7,0^{+}){\to}\bar{D}_{s}\bar{B}_{c}\right]}
\sim 2.6\,.
\end{equation}
For the highest axial-vector states, we have
\begin{equation}
\Gamma_{\bar{D}^{*}\bar{B}_{c}^{*}}:\Gamma_{\bar{D}^{*}\bar{B}_{c}}:\Gamma_{\bar{D}\bar{B}_{c}^{*}}
\sim 23.8:10.7:1
\end{equation}
for $T(nb\bar{c}\bar{c},8454.5,1^{+})$ and
\begin{equation}
\Gamma_{\bar{D}_{s}^{*}\bar{B}_{c}^{*}}:\Gamma_{\bar{D}_{s}^{*}\bar{B}_{c}}:\Gamma_{\bar{D}_{s}\bar{B}_{c}^{*}}
\sim 15.7:5.1:1
\end{equation}
for $T(sb\bar{c}\bar{c},8553.3,1^{+})$.
The $\bar{D}_{(s)}^{*}\bar{B}_{c}^{*}$ and
$\bar{D}_{(s)}^{*}\bar{B}_{c}$ modes are more important.
However, for the second higher axial-vector states of
$qb\bar{c}\bar{c}$ tetraquarks, the partial decay widths of
$\bar{D}_{(s)}^{*}\bar{B}_{c}^{*}$, $\bar{D}_{(s)}^{*}\bar{B}_{c}$
and $\bar{D}_{(s)}\bar{B}_{c}^{*}$ modes are comparable, though the
first mode is relatively smaller.
%

\subsection{The $qb\bar{b}\bar{b}$ and $qc\bar{b}\bar{b}$ systems}
\label{Sec:qbbb+qcbb}

\begin{figure*}
    \begin{tabular}{ccc}
        \includegraphics[width=450pt]{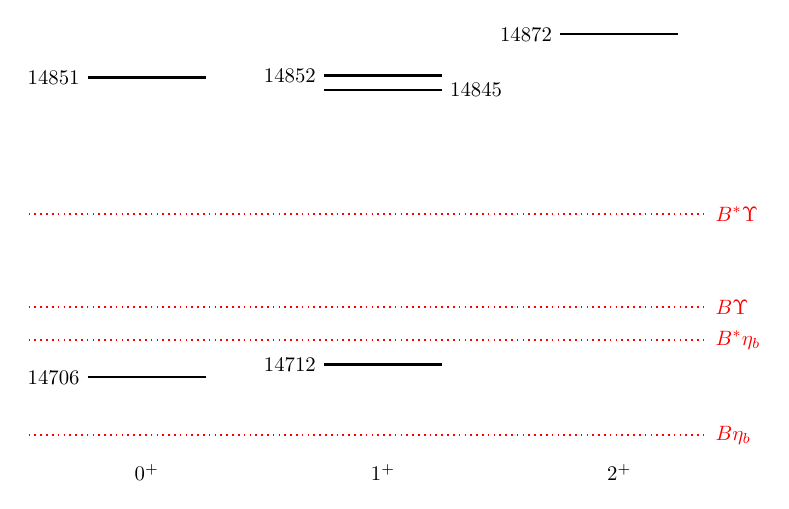}\\
        (a) $nb\bar{b}\bar{b}$ states\\
        &&\\
        \includegraphics[width=450pt]{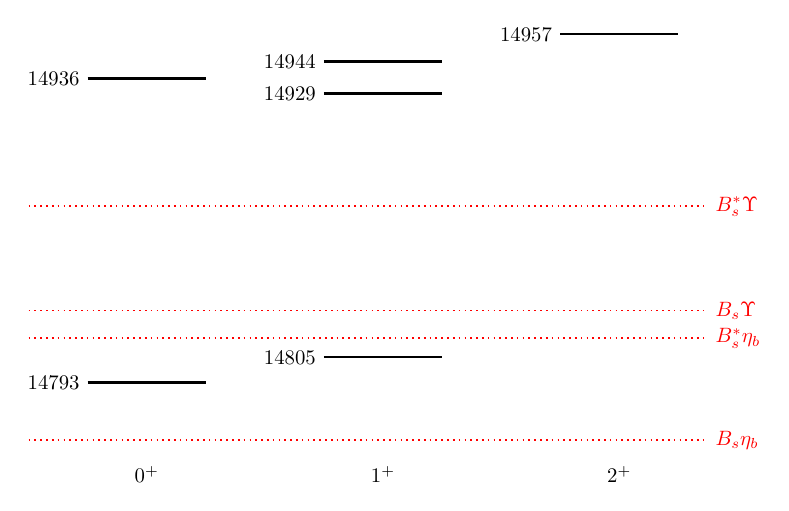}\\
        (b) $sb\bar{b}\bar{b}$ states\\
    \end{tabular}
    \caption{Mass spectra of the $nb\bar{b}\bar{b}$ and $sb\bar{b}\bar{b}$ tetraquark states. The dotted lines indicate various meson-meson thresholds. The masses are all in units of MeV.}
    \label{fig:nbbb+sbbb}
\end{figure*}
%
\begin{figure*}
    \begin{tabular}{ccc}
        \includegraphics[width=450pt]{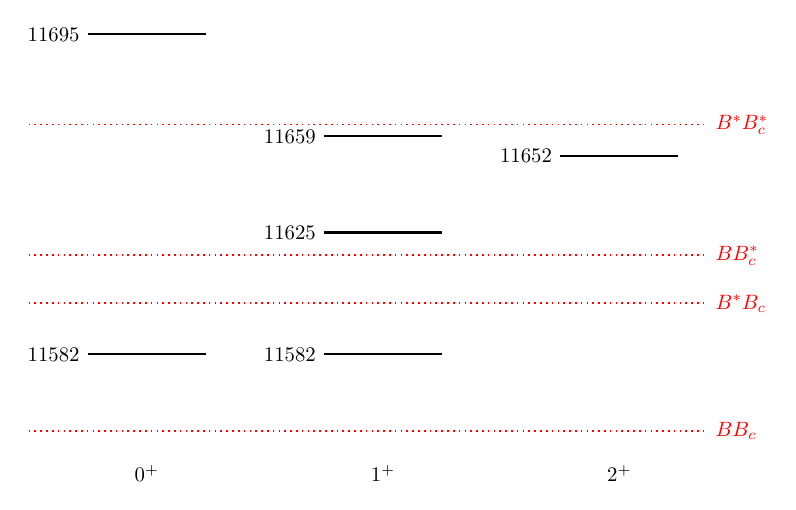}\\
        (a) $nc\bar{b}\bar{b}$ states\\
        &&\\
        \includegraphics[width=450pt]{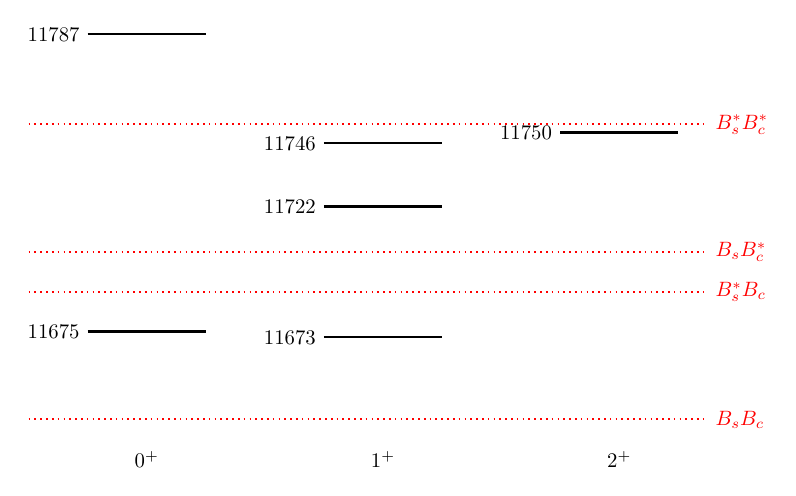}\\
        (b) $sc\bar{b}\bar{b}$ states\\
    \end{tabular}
    \caption{Mass spectra of the $nc\bar{b}\bar{b}$ and $sc\bar{b}\bar{b}$ tetraquark states. The dotted lines indicate various meson-meson thresholds. Here the predicted mass $M_{B_{c}^{*}}=6338~\text{MeV}$ of Godfrey {\it et~al.}~\cite{Godfrey:1985xj} is used. The masses are all in units of MeV.}
    \label{fig:ncbb+scbb}
\end{figure*}
%

\begin{table}
    \centering
    \caption{Masses and eigenvectors of the $nb\bar{b}\bar{b}$ and $sb\bar{b}\bar{b}$ tetraquarks. The masses are all in units of MeV.}
    \label{table:mass:nbbb+sbbb}
    \begin{tabular}{ccccccc}
        \toprule[1pt]
        \toprule[1pt]
        System&$J^{P}$&Mass&Eigenvector\\
        \midrule[1pt]
        $nb\bar{b}\bar{b}$&$0^{+}$
        &$14706.1$&$\{0.932,0.362\}$\\
        &
        &$14850.9$&$\{-0.362,0.932\}$\\
        &$1^{+}$
        &$14712.1$&$\{0.977,0.049,-0.208\}$\\
        &
        &$14845.1$&$\{-0.097,0.969,-0.228\}$\\
        &
        &$14851.6$&$\{0.191,0.243,0.951\}$\\
        &$2^{+}$
        &$14871.8$&$\{1\}$\\
        \midrule[1pt]
        $sb\bar{b}\bar{b}$&$0^{+}$
        &$14793.1$&$\{0.924,0.382\}$\\
        &
        &$14936.2$&$\{-0.382,0.924\}$\\
        &$1^{+}$
        &$14804.5$&$\{0.976,0.041,-0.215\}$\\
        &
        &$14929.2$&$\{0.058,-0.996,0.070\}$\\
        &
        &$14943.7$&$\{-0.211,-0.081,-0.974\}$\\
        &$2^{+}$
        &$14956.6$&$\{1\}$\\
        \bottomrule[1pt]
        \bottomrule[1pt]
    \end{tabular}
\end{table}
%
\begin{table}
    \centering
    \caption{Masses and eigenvectors of the $nc\bar{b}\bar{b}$ and $sc\bar{b}\bar{b}$ tetraquarks. The masses are all in units of MeV.}
    \label{table:mass:ncbb+scbb}
    \begin{tabular}{ccccccc}
        \toprule[1pt]
        \toprule[1pt]
        System&$J^{P}$&Mass&Eigenvector\\
        \midrule[1pt]
        $nc\bar{b}\bar{b}$&$0^{+}$
        &$11581.7$&$\{0.510,0.860\}$\\
        &
        &$11694.9$&$\{0.860,-0.510\}$\\
        &$1^{+}$
        &$11582.2$&$\{-0.425,-0.133,0.895\}$\\
        &
        &$11624.7$&$\{-0.070,-0.981,-0.179\}$\\
        &
        &$11658.6$&$\{-0.902,0.139,-0.408\}$\\
        &$2^{+}$
        &$11651.7$&$\{1\}$\\
        \midrule[1pt]
        $sc\bar{b}\bar{b}$&$0^{+}$
        &$11675.1$&$\{0.545,0.838\}$\\
        &
        &$11787.2$&$\{0.838,-0.545\}$\\
        &$1^{+}$
        &$11673.3$&$\{-0.465,-0.096,0.880\}$\\
        &
        &$11722.2$&$\{-0.089,-0.984,-0.155\}$\\
        &
        &$11746.3$&$\{0.881,-0.150,0.449\}$\\
        &$2^{+}$
        &$11750.2$&$\{1\}$\\
        \bottomrule[1pt]
        \bottomrule[1pt]
    \end{tabular}
\end{table}
%

\begin{table}
    \centering
    \caption{The eigenvectors of the $nb\bar{b}\bar{b}$ tetraquarks in the $n\bar{b}{\otimes}b\bar{b}$ configuration. The masses are all in units of MeV.}
    \label{table:eigenvector:nbbb}
    \begin{tabular}{ccccccc}
        \toprule[1pt]
        \toprule[1pt]
        System&$J^{P}$&Mass&${B}^{*}{\Upsilon}$&${B}^{*}{\eta_{b}}$&${B}{\Upsilon}$&${B}{\eta_{b}}$\\
        \midrule[1pt]
        $nb\bar{b}\bar{b}$&$0^{+}$
        &$14706.1$&$0.554$&&&$0.562$\\
        &
        &$14850.9$&$-0.525$&&&$0.318$\\
        &$1^{+}$
        &$14712.1$&$0.479$&$0.479$&$-0.439$\\
        &
        &$14845.1$&$-0.149$&$0.422$&$0.369$\\
        &
        &$14851.6$&$0.498$&$-0.098$&$0.296$\\
        &$2^{+}$
        &$14871.8$&$0.577$\\
        \bottomrule[1pt]
        \bottomrule[1pt]
    \end{tabular}
\end{table}
%
\begin{table}
    \centering
    \caption{The eigenvectors of the $sb\bar{b}\bar{b}$ tetraquarks in the $s\bar{b}{\otimes}b\bar{b}$ configuration. The masses are all in units of MeV.}
    \label{table:eigenvector:sbbb}
    \begin{tabular}{ccccccc}
        \toprule[1pt]
        \toprule[1pt]
        System&$J^{P}$&Mass&${B}_{s}^{*}{\Upsilon}$&${B}_{s}^{*}{\eta_{b}}$&${B}_{s}{\Upsilon}$&${B}_{s}{\eta_{b}}$\\
        \midrule[1pt]
        $sb\bar{b}\bar{b}$&$0^{+}$
        &$14793.1$&$0.543$&&&$0.568$\\
        &
        &$14936.2$&$-0.537$&&&$0.306$\\
        &$1^{+}$
        &$14804.5$&$0.476$&$0.477$&$-0.444$\\
        &
        &$14929.2$&$0.062$&$-0.403$&$-0.410$\\
        &
        &$14943.7$&$-0.520$&$0.162$&$-0.228$\\
        &$2^{+}$
        &$14956.6$&$0.577$\\
        \bottomrule[1pt]
        \bottomrule[1pt]
    \end{tabular}
\end{table}
%
\begin{table}
    \centering
    \caption{The values of $k\cdot|c_{i}|^2$ for the $nb\bar{b}\bar{b}$ tetraquarks (in unit of MeV).}
    \label{table:kc_i^2:nbbb}
    \begin{tabular}{ccccccccccccc}
        \toprule[1pt]
        \toprule[1pt]
        System&$J^{P}$&Mass&${B}^{*}{\Upsilon}$&${B}^{*}{\eta_{b}}$&${B}{\Upsilon}$&${B}{\eta_{b}}$\\
        \midrule[1pt]
        $nb\bar{b}\bar{b}$&$0^{+}$
        &$14706.1$&$\times$&&&$136.4$\\
        &
        &$14850.9$&$185.2$&&&$109.7$\\
        &$1^{+}$
        &$14712.1$&$\times$&$\times$&$\times$\\
        &
        &$14845.1$&$14.2$&$162.0$&$115.6$\\
        &
        &$14851.6$&$167.7$&$8.9$&$76.4$\\
        &$2^{+}$
        &$14871.8$&$257.0$\\
        \bottomrule[1pt]
        \bottomrule[1pt]
    \end{tabular}
\end{table}
%
\begin{table}
    \centering
    \caption{The values of $k\cdot|c_{i}|^2$ for the $sb\bar{b}\bar{b}$ tetraquarks (in unit of MeV).}
    \label{table:kc_i^2:sbbb}
    \begin{tabular}{ccccccccccccc}
        \toprule[1pt]
        \toprule[1pt]
        System&$J^{P}$&Mass&${B}_{s}^{*}{\Upsilon}$&${B}_{s}^{*}{\eta_{b}}$&${B}_{s}{\Upsilon}$&${B}_{s}{\eta_{b}}$\\
        \midrule[1pt]
        $sb\bar{b}\bar{b}$&$0^{+}$
        &$14793.1$&$\times$&&&$139.5$\\
        &
        &$14936.2$&$186.2$&&&$101.6$\\
        &$1^{+}$
        &$14804.5$&$\times$&$\times$&$\times$\\
        &
        &$14929.2$&$2.3$&$144.9$&$140.7$\\
        &
        &$14943.7$&$185.0$&$24.8$&$46.6$\\
        &$2^{+}$
        &$14956.5$&$249.2$\\
        \bottomrule[1pt]
        \bottomrule[1pt]
    \end{tabular}
\end{table}
%
\begin{table}
    \centering
    \caption{The partial width ratios for the $nb\bar{b}\bar{b}$ tetraquarks. For each state, we choose one mode as the reference channel, and the partial width ratios of the other channels are calculated relative to this channel. The masses are all in unit of MeV.}
    \label{table:R:nbbb}
    \begin{tabular}{ccccccccccccc}
        \toprule[1pt]
        \toprule[1pt]
        System&$J^{P}$&Mass&${B}^{*}{\Upsilon}$&${B}^{*}{\eta_{b}}$&${B}{\Upsilon}$&${B}{\eta_{b}}$\\
        \midrule[1pt]
        $nb\bar{b}\bar{b}$&$0^{+}$
        &$14706.1$&$\times$&&&$1$\\
        &
        &$14850.9$&$1.7$&&&$1$\\
        &$1^{+}$
        &$14712.1$&$\times$&$\times$&$\times$\\
        &
        &$14845.1$&$0.1$&$1.4$&$1$\\
        &
        &$14851.6$&$2.2$&$0.1$&$1$\\
        &$2^{+}$
        &$14871.8$&$1$\\
        \bottomrule[1pt]
        \bottomrule[1pt]
    \end{tabular}
\end{table}
%
\begin{table}
    \centering
    \caption{The partial width ratios for the $sb\bar{b}\bar{b}$ tetraquarks. For each state, we choose one mode as the reference channel, and the partial width ratios of the other channels are calculated relative to this channel. The masses are all in unit of MeV.}
    \label{table:R:sbbb}
    \begin{tabular}{ccccccccccccc}
        \toprule[1pt]
        \toprule[1pt]
        System&$J^{P}$&Mass&${B}^{*}{\Upsilon}$&${B}^{*}{\eta_{b}}$&${B}{\Upsilon}$&${B}{\eta_{b}}$\\
        \midrule[1pt]
        $sb\bar{b}\bar{b}$&$0^{+}$
        &$14793.1$&$\times$&&&$1$\\
        &
        &$14936.2$&$1.8$&&&$1$\\
        &$1^{+}$
        &$14804.5$&$\times$&$\times$&$\times$\\
        &
        &$14929.2$&$0.02$&$1.03$&$1$\\
        &
        &$14943.7$&$3.97$&$0.53$&$1$\\
        &$2^{+}$
        &$14956.5$&$1$\\
        \bottomrule[1pt]
        \bottomrule[1pt]
    \end{tabular}
\end{table}
%

\begin{table}
    \centering
    \caption{The eigenvectors of the $nc\bar{b}\bar{b}$ tetraquarks in the $n\bar{b}{\otimes}c\bar{b}$ configuration. The masses are all in units of MeV.}
    \label{table:eigenvector:ncbb}
    \begin{tabular}{ccccccc}
        \toprule[1pt]
        \toprule[1pt]
        System&$J^{P}$&Mass&${B}^{*}{B}_{c}^{*}$&${B}^{*}{B}_{c}$&${B}{B}_{c}^{*}$&${B}{B}_{c}$\\
        \midrule[1pt]
        $nc\bar{b}\bar{b}$&$0^{+}$
        &$11581.7$&$0.112$&&&$0.638$\\
        &
        &$11694.9$&$0.756$&&&$0.096$\\
        &$1^{+}$
        &$11582.2$&$0.120$&$-0.486$&$0.378$\\
        &
        &$11624.7$&$-0.114$&$-0.378$&$-0.424$\\
        &
        &$11658.6$&$-0.688$&$-0.194$&$0.307$\\
        &$2^{+}$
        &$11651.7$&$0.577$\\
        \bottomrule[1pt]
        \bottomrule[1pt]
    \end{tabular}
\end{table}
%
\begin{table}
    \centering
    \caption{The eigenvectors of the $sc\bar{b}\bar{b}$ tetraquarks in the $s\bar{b}{\otimes}c\bar{b}$ configuration. The masses are all in units of MeV.}
    \label{table:eigenvector:scbb}
    \begin{tabular}{ccccccc}
        \toprule[1pt]
        \toprule[1pt]
        System&$J^{P}$&Mass&${B}_{s}^{*}{B}_{c}^{*}$&${B}_{s}^{*}{B}_{c}$&${B}_{s}{B}_{c}^{*}$&${B}_{s}{B}_{c}$\\
        \midrule[1pt]
        $sc\bar{b}\bar{b}$&$0^{+}$
        &$11675.1$&$0.143$&&&$0.642$\\
        &
        &$11787.2$&$0.750$&&&$0.070$\\
        &$1^{+}$
        &$11673.3$&$0.091$&$-0.483$&$0.404$\\
        &
        &$11722.2$&$-0.114$&$-0.393$&$-0.410$\\
        &
        &$11746.3$&$0.692$&$0.169$&$-0.291$\\
        &$2^{+}$
        &$11750.2$&$0.577$\\
        \bottomrule[1pt]
        \bottomrule[1pt]
    \end{tabular}
\end{table}
%
\begin{table}
    \centering
    \caption{The values of $k\cdot|c_{i}|^2$ for the $nc\bar{b}\bar{b}$ tetraquarks (in unit of MeV).}
    \label{table:kc_i^2:ncbb}
    \begin{tabular}{ccccccccccccc}
        \toprule[1pt]
        \toprule[1pt]
        System&$J^{P}$&Mass&${B}^{*}{B}_{c}^{*}$&${B}^{*}{B}_{c}$&${B}{B}_{c}^{*}$&${B}{B}_{c}$\\
        \midrule[1pt]
        $nc\bar{b}\bar{b}$&$0^{+}$
        &$11581.7$&$\times$&&&$160.9$\\
        &
        &$11694.9$&$246.9$&&&$8.4$\\
        &$1^{+}$
        &$11582.2$&$\times$&$\times$&$\times$\\
        &
        &$11624.7$&$\times$&$54.2$&$36.7$\\
        &
        &$11658.6$&$\times$&$21.9$&$46.0$\\
        &$2^{+}$
        &$11651.7$&$\times$\\
        \bottomrule[1pt]
        \bottomrule[1pt]
    \end{tabular}
\end{table}
%
\begin{table}
    \centering
    \caption{The values of $k\cdot|c_{i}|^2$ for the $sc\bar{b}\bar{b}$ tetraquarks (in unit of MeV).}
    \label{table:kc_i^2:scbb}
    \begin{tabular}{ccccccccccccc}
        \toprule[1pt]
        \toprule[1pt]
        System&$J^{P}$&Mass&${B}_{s}^{*}{B}_{c}^{*}$&${B}_{s}^{*}{B}_{c}$&${B}_{s}{B}_{c}^{*}$&${B}_{s}{B}_{c}$\\
        \midrule[1pt]
        $sc\bar{b}\bar{b}$&$0^{+}$
        &$11675.1$&$\times$&&&$180.6$\\
        &
        &$11787.2$&$250.2$&&&$4.5$\\
        &$1^{+}$
        &$11673.3$&$\times$&$\times$&$\times$\\
        &
        &$11722.2$&$\times$&$66.5$&$53.6$\\
        &
        &$11746.3$&$\times$&$16.3$&$41.7$\\
        &$2^{+}$
        &$11750.2$&$\times$\\
        \bottomrule[1pt]
        \bottomrule[1pt]
    \end{tabular}
\end{table}
%
\begin{table}
    \centering
    \caption{The partial width ratios for the $nc\bar{b}\bar{b}$ tetraquarks. For each state, we choose one mode as the reference channel, and the partial width ratios of the other channels are calculated relative to this channel. The masses are all in unit of MeV.}
    \label{table:R:ncbb}
    \begin{tabular}{ccccccccccccc}
        \toprule[1pt]
        \toprule[1pt]
        System&$J^{P}$&Mass&${B}^{*}{B}_{c}^{*}$&${B}^{*}{B}_{c}$&${B}{B}_{c}^{*}$&${B}{B}_{c}$\\
        \midrule[1pt]
        $nc\bar{b}\bar{b}$&$0^{+}$
        &$11581.7$&$\times$&&&$1$\\
        &
        &$11694.9$&$29.5$&&&$1$\\
        &$1^{+}$
        &$11582.2$&$\times$&$\times$&$\times$\\
        &
        &$11624.7$&$\times$&$1.5$&$1$\\
        &
        &$11658.6$&$\times$&$0.5$&$1$\\
        &$2^{+}$
        &$11651.7$&$\times$\\
        \bottomrule[1pt]
        \bottomrule[1pt]
    \end{tabular}
\end{table}
%
\begin{table}
    \centering
    \caption{The partial width ratios for the $sc\bar{b}\bar{b}$ tetraquarks. For each state, we choose one mode as the reference channel, and the partial width ratios of the other channels are calculated relative to this channel. The masses are all in unit of MeV.}
    \label{table:R:scbb}
    \begin{tabular}{ccccccccccccc}
        \toprule[1pt]
        \toprule[1pt]
        System&$J^{P}$&Mass&${B}_{s}^{*}{B}_{c}^{*}$&${B}_{s}^{*}{B}_{c}$&${B}_{s}{B}_{c}^{*}$&${B}_{s}{B}_{c}$\\
        \midrule[1pt]
        $sc\bar{b}\bar{b}$&$0^{+}$
        &$11675.1$&$\times$&&&$1$\\
        &
        &$11787.2$&$55.7$&&&$1$\\
        &$1^{+}$
        &$11673.3$&$\times$&$\times$&$\times$\\
        &
        &$11722.2$&$\times$&$1.2$&$1$\\
        &
        &$11746.3$&$\times$&$0.4$&$1$\\
        &$2^{+}$
        &$11750.2$&$\times$\\
        \bottomrule[1pt]
        \bottomrule[1pt]
    \end{tabular}
\end{table}

Next we consider the $qb\bar{b}\bar{b}$ and $qc\bar{b}\bar{b}$
tetraquarks.
%
%
Their masses and eigenvectors are listed in
Tables~\ref{table:mass:nbbb+sbbb}--\ref{table:mass:ncbb+scbb}.
Their relative position are ploted in Figs.~\ref{fig:nbbb+sbbb}--\ref{fig:ncbb+scbb}.
By replacing the three charm quark/antiquarks of the
$qc\bar{c}\bar{c}$ tetraquarks into bottom quark/antiquarks, we can
obtain the $qb\bar{b}\bar{b}$ tetraquarks.
Thus we expect that the two systems share some similar properties.
Specifically, the quantum numbers of the lowest and highest
$qb\bar{b}\bar{b}$ tetraquarks are $J^{P}=0^{+}$ and $J^{P}=2^{+}$
respectively.
Their ground states are $T(nb\bar{b}\bar{b},14706.1,0^{+})$ and
$T(sb\bar{b}\bar{b},14793.1,0^{+})$, whose dominant color
configurations are $6_{c}\otimes\bar{6}_{c}$ ($>85\%$).
The color triplet component is mostly in the higher scalar states.
The axial-vector states are similar. The $6_{c}\otimes\bar{6}_{c}$
component is mostly in the lightest states ($>90\%$), and the
$\bar{3}_{c}\otimes3_{c}$ components are mostly in the higher
states.
Just like the $qc\bar{c}\bar{c}$ cases, the underlining reason is
the color interaction
\begin{equation}
\Braket{H_{\text{C}}\left(qb\bar{b}\bar{b}\right)}
= m_{q\bar{b}} + m_{b\bar{b}} - \frac{3}{2} \delta{m}'
\Braket{V_{12}^{\text{C}}+V_{34}^{\text{C}}}
\end{equation}
where
\begin{equation}
\delta{m}' = \frac{m_{qb}+m_{bb}-m_{q\bar{b}}-m_{b\bar{b}}}{4} \sim
45~\text{MeV}\,.
\end{equation}
Note that the $\Braket{V_{12}^{\text{C}}+V_{34}^{\text{C}}}$ is
diagonal in the $6_{c}\otimes\bar{6}_{c}$ and
$\bar{3}_{c}\otimes3_{c}$ color bases, with matrix elements $2/3$
and $-4/3$ respectively.
Thus the $6_{c}\otimes\bar{6}_{c}$ configuration is favored.

The $qc\bar{b}\bar{b}$ tetraquarks turn out to be quite different
from the $qb\bar{b}\bar{b}$ tetraquarks.
The quantum number of the highest states becomes $J^{P}=0^{+}$.
For the $nc\bar{b}\bar{b}$ tetraquarks, the lightest axial-vector
state lies only $0.5~\text{MeV}$ above the lightest scalar state.
And the ground state of the $sc\bar{b}\bar{b}$ tetraquarks has
quantum number $J^{P}=1^{+}$.
Moreover, the dominant color configuration of the lower mass states
becomes $\bar{3}_{c}\otimes3_{c}$.
For example, the $T(nc\bar{b}\bar{b},11581.7,0^{+})$ has $74\%$ of
$\bar{3}_{c}\otimes3_{c}$ component, and the
$T(sc\bar{b}\bar{b},11675.1,0^{+})$ has $70\%$.
We again resort to the color interaction
\begin{equation}
\Braket{H_{\text{C}}\left(qc\bar{b}\bar{b}\right)}
= m_{q\bar{b}} + m_{c\bar{b}} - \frac{3}{2} \delta{m}''
\Braket{V_{12}^{\text{C}}+V_{34}^{\text{C}}}
\end{equation}
where
\begin{align}
\delta{m}''
= \frac{m_{qc}+m_{bb}-m_{q\bar{b}}-m_{c\bar{b}}}{4}\,.
\end{align}
Here $\delta{m}''=-6.5~\text{MeV}$ for $q=n$, and
$\delta{m}''=-3.1~\text{MeV}$ for $q=s$.
The small negative values indicate that the
$\bar{3}_{c}\otimes3_{c}$ configuration has slightly lower mass.
The CM interaction
\begin{align}
&\Braket{H_{\text{CM}}\left(qc\bar{b}\bar{b}\right)}
\notag\\
={}&
\begin{pmatrix}
\frac{1}{4}\left(v_{qc}+v_{bb}\right)&-\frac{\sqrt{6}}{4}\left(v_{q\bar{b}}+v_{c\bar{b}}\right)\\
-\frac{\sqrt{6}}{4}\left(v_{q\bar{b}}+v_{c\bar{b}}\right)&\frac{1}{6}\left(v_{qc}+v_{bb}\right)-\frac{1}{3}\left(v_{q\bar{b}}+v_{c\bar{b}}\right)\\
\end{pmatrix}
\end{align}
also favors the $\bar{3}_{c}\otimes3_{c}$ configuration.
Thus for both the scalar and the axial-vector states, the lower mass
states are dominated by the $\bar{3}_{c}\otimes3_{c}$ components,
while the higher mass states have more $6_{c}\otimes\bar{6}_{c}$
components.

Before concluding this section, we would like to compare the present
results with that of Refs.~\cite{Weng:2020jao,Weng:2021hje}.
It is interesting to note that the $qb\bar{b}\bar{b}$ tetraquarks
(as well as the $qc\bar{c}\bar{c}$ and $qb\bar{c}\bar{c}$
tetraquarks) are very similar to the fully heavy $cb\bar{b}\bar{b}$
tetraquarks, while the $qc\bar{b}\bar{b}$ tetraquarks are more like
the doubly heavy tetraquarks.
This is quite natural since that the triply heavy tetraquarks are
intermediate states between the doubly and fully heavy tetraquarks.
The reason is $m_{q}{\ll}m_{c}{\ll}m_{b}$.
A detailed dynamical study of the dependence of the spectrum and
wave function with respect to the quark masses would be very
important for decoding the nature of
tetraquarks~\cite{Richard:2019cmi,Richard:2021lce}.

Next we consider their decay properties.
We transform the eigenvectors of the $qb\bar{b}\bar{b}$
($qc\bar{b}\bar{b}$) tetraquarks into the
$q\bar{b}{\otimes}b\bar{b}$ ($q\bar{b}{\otimes}c\bar{b}$)
configuration, and calculate the values of $k\cdot|c_i|^2$ and
partial width ratios of all possible decay channels.
The corresponding results are listed in
Tables~\ref{table:eigenvector:nbbb}--\ref{table:R:scbb}.
We find six states which cannot decay into two mesons through
$S$-wave.
They are $T(nb\bar{b}\bar{b},14712.1,1^{+})$,
$T(sb\bar{b}\bar{b},14804.5,1^{+})$,
$T(nc\bar{b}\bar{b},11582.2,1^{+})$,
$T(nc\bar{b}\bar{b},11651.7,2^{+})$,
$T(sc\bar{b}\bar{b},11673.3,1^{+})$ and
$T(sc\bar{b}\bar{b},11750.2,2^{+})$.
Because of the conservation of the angular momentum and parity,
their decays into the thresholds below them must be of higher wave
($D$-wave for example), which are highly suppressed.
Thus we expect them to be narrow.

However, we need to bear in mind that our model is an oversimplified
one.
The dynamical effects are not included explicitly, but rather
concealed in the interaction strengths $\{m_{ij},v_{ij}\}$.
In the present work, we fit them from the conventional mesons and
baryons.
But they may be altered in the tetraquarks because of their wave
function dependence.
Nonetheless, even if the masses of these states are pushed upward
above their $S$-wave channels, their phase spaces are still
relatively small.
Thus their decay widths are still smaller compared to those of other
states.
A dynamical calculation may help settle this issue.
We hope the future experiments can search for these states.
%

\subsection{The $qc\bar{c}\bar{b}$ and $qb\bar{c}\bar{b}$ systems}
\label{Sec:qccb+qbcb}

\begin{figure*}
    \begin{tabular}{ccc}
        \includegraphics[width=450pt]{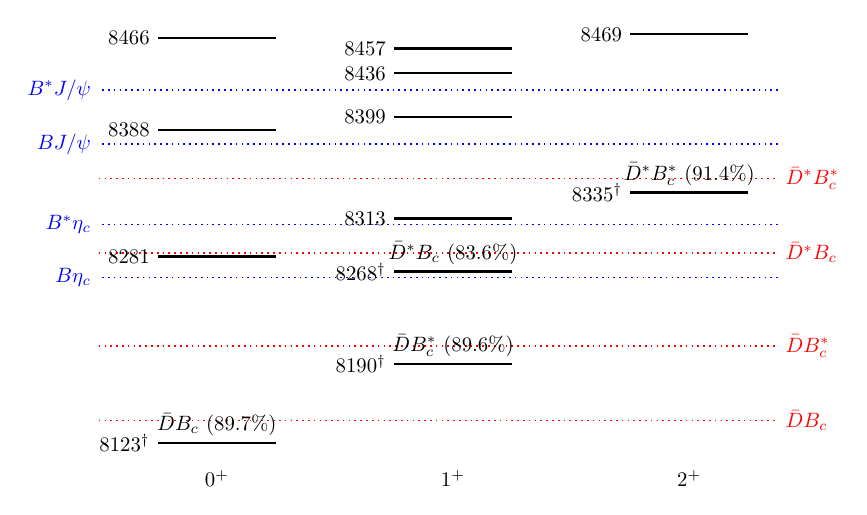}\\
        (a) $nc\bar{c}\bar{b}$ states\\
        &&\\
        \includegraphics[width=450pt]{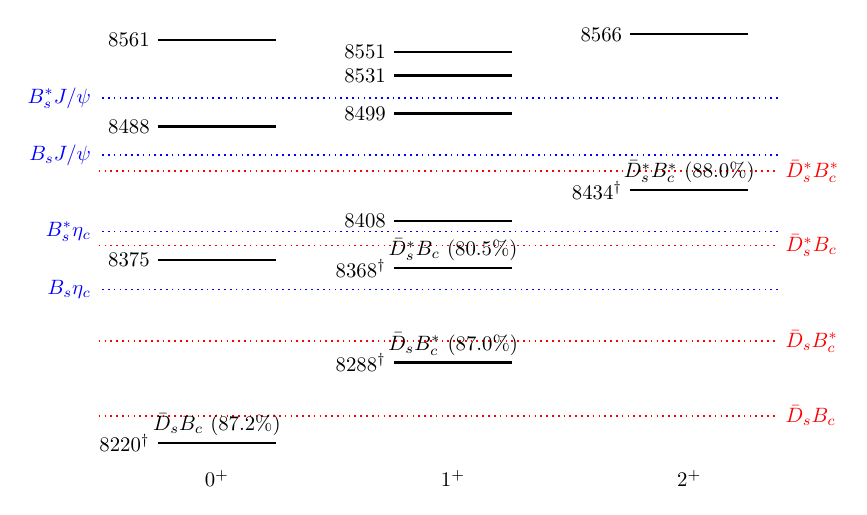}\\
        (b) $sc\bar{c}\bar{b}$ states\\
    \end{tabular}
    \caption{Mass spectra of the $nc\bar{c}\bar{b}$ and $sc\bar{c}\bar{b}$ tetraquark states. The dotted lines indicate various meson-meson thresholds. The scattering states are marked with a dagger ($\dagger$), along with the proportion of their dominant components. Here the predicted mass $M_{B_{c}^{*}}=6338~\text{MeV}$ of Godfrey {\it et~al.}~\cite{Godfrey:1985xj} is used. The masses are all in units of MeV.}
    \label{fig:nccb+sccb}
\end{figure*}
%
\begin{figure*}
    \begin{tabular}{ccc}
        \includegraphics[width=450pt]{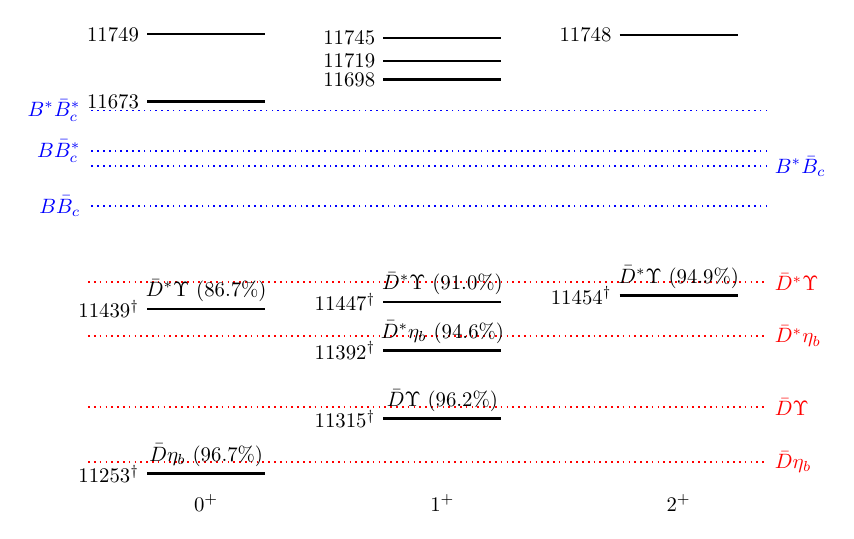}\\
        (a) $nb\bar{c}\bar{b}$ states\\
        &&\\
        \includegraphics[width=450pt]{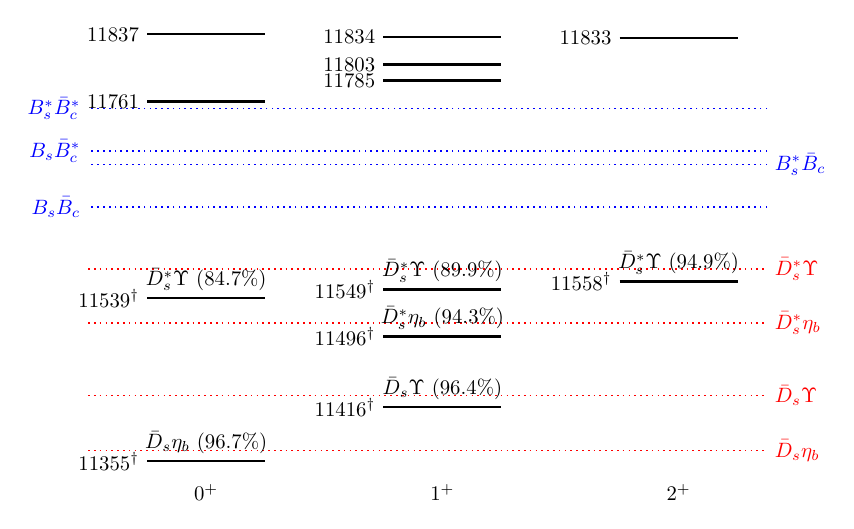}\\
        (b) $sb\bar{c}\bar{b}$ states\\
    \end{tabular}
    \caption{Mass spectra of the $nb\bar{c}\bar{b}$ and $sb\bar{c}\bar{b}$ tetraquark states. The dotted lines indicate various meson-meson thresholds. The scattering states are marked with a dagger ($\dagger$), along with the proportion of their dominant components. Here the predicted mass $M_{B_{c}^{*}}=6338~\text{MeV}$ of Godfrey {\it et~al.}~\cite{Godfrey:1985xj} is used. The masses are all in units of MeV.}
    \label{fig:nbcb+sbcb}
\end{figure*}
%

\begin{table*}
    \centering
    \caption{Masses and eigenvectors of the $nc\bar{c}\bar{b}$ and $sc\bar{c}\bar{b}$ tetraquarks. The masses are all in units of MeV.}
    \label{table:mass:nccb+sccb}
    \begin{tabular}{ccccccc}
        \toprule[1pt]
        \toprule[1pt]
        System&$J^{P}$&Mass&Eigenvector&Scattering~state\\
        \midrule[1pt]
        $nc\bar{c}\bar{b}$&$0^{+}$
        &$8123.2$&$\{0.864,0.246,0.273,0.343\}$&$\bar{D}{B}_{c}$~($89.7\%$)\\
        &
        &$8281.2$&$\{-0.394,0.812,0.425,0.073\}$\\
        &
        &$8388.0$&$\{-0.199,0.091,-0.502,0.837\}$\\
        &
        &$8466.4$&$\{-0.240,-0.522,0.702,0.421\}$\\
        &$1^{+}$
        &$8190.4$&$\{0.770,-0.430,0.259,0.229,-0.169,0.271\}$&$\bar{D}{B}_{c}^{*}$~($89.6\%$)\\
        &
        &$8268.0$&$\{-0.562,-0.380,0.609,-0.051,-0.317,0.256\}$&$\bar{D}^{*}{B}_{c}$~($83.6\%$)\\
        &
        &$8312.6$&$\{0.142,0.732,0.623,0.235,0.003,-0.033\}$\\
        &
        &$8398.9$&$\{0.025,0.123,0.029,-0.379,0.530,0.747\}$\\
        &
        &$8435.7$&$\{-0.264,-0.081,-0.159,0.861,0.315,0.242\}$\\
        &
        &$8457.3$&$\{-0.030,0.336,-0.385,0.070,-0.700,0.493\}$\\
        &$2^{+}$
        &$8335.1$&$\{0.950,0.313\}$&$\bar{D}^{*}{B}_{c}^{*}$~($91.4\%$)\\
        &
        &$8468.9$&$\{-0.313,0.950\}$\\
        \midrule[1pt]
        $sc\bar{c}\bar{b}$&$0^{+}$
        &$8220.1$&$\{0.880,0.223,0.243,0.343\}$&$\bar{D}_{s}{B}_{c}$~($87.2\%$)\\
        &
        &$8375.2$&$\{0.351,-0.817,-0.456,-0.046\}$\\
        &
        &$8488.1$&$\{0.238,-0.092,0.437,-0.862\}$\\
        &
        &$8561.0$&$\{-0.215,-0.524,0.736,0.370\}$\\
        &$1^{+}$
        &$8287.8$&$\{0.789,-0.431,0.240,0.204,-0.152,0.264\}$&$\bar{D}_{s}{B}_{c}^{*}$~($87.0\%$)\\
        &
        &$8368.4$&$\{-0.545,-0.397,0.623,-0.013,-0.309,0.249\}$&$\bar{D}_{s}^{*}{B}_{c}$~($80.5\%$)\\
        &
        &$8407.5$&$\{-0.168,-0.727,-0.622,-0.220,0.049,0.079\}$\\
        &
        &$8498.8$&$\{0.006,0.183,0.043,-0.339,0.487,0.783\}$\\
        &
        &$8531.2$&$\{0.227,0.033,0.164,-0.890,-0.266,-0.239\}$\\
        &
        &$8551.1$&$\{-0.023,0.307,-0.373,0.045,-0.756,0.439\}$\\
        &$2^{+}$
        &$8434.2$&$\{0.966,0.259\}$&$\bar{D}_{s}^{*}{B}_{c}^{*}$~($88.0\%$)\\
        &
        &$8566.3$&$\{-0.259,0.966\}$\\
        \bottomrule[1pt]
        \bottomrule[1pt]
    \end{tabular}
\end{table*}
%
\begin{table*}
    \centering
    \caption{Masses and eigenvectors of the $nb\bar{c}\bar{b}$ and $sb\bar{c}\bar{b}$ tetraquarks. The masses are all in units of MeV.}
    \label{table:mass:nbcb+sbcb}
    \begin{tabular}{ccccccc}
        \toprule[1pt]
        \toprule[1pt]
        System&$J^{P}$&Mass&Eigenvector&Scattering~state\\
        \midrule[1pt]
        $nb\bar{c}\bar{b}$&$0^{+}$
        &$11253.4$&$\{0.816,0.365,0.364,0.261\}$&$\bar{D}{\eta}_{b}$~($96.7\%$)\\
        &
        &$11438.6$&$\{0.463,-0.835,0.011,-0.296\}$&$\bar{D}^{*}{\Upsilon}$~($86.7\%$)\\
        &
        &$11673.1$&$\{0.160,-0.165,-0.688,0.688\}$\\
        &
        &$11748.7$&$\{0.306,0.377,-0.628,-0.609\}$\\
        &$1^{+}$
        &$11314.8$&$\{0.702,-0.408,0.397,0.298,-0.215,0.219\}$&$\bar{D}{\Upsilon}$~($96.2\%$)\\
        &
        &$11392.4$&$\{-0.609,-0.450,0.515,-0.218,-0.240,0.239\}$&$\bar{D}^{*}{\eta}_{b}$~($94.6\%$)\\
        &
        &$11446.6$&$\{-0.024,-0.688,-0.655,-0.028,-0.231,-0.205\}$&$\bar{D}^{*}{\Upsilon}$~($91.0\%$)\\
        &
        &$11697.5$&$\{0.080,-0.232,-0.184,-0.236,0.675,0.628\}$\\
        &
        &$11719.1$&$\{0.353,0.101,0.028,-0.889,-0.267,-0.047\}$\\
        &
        &$11744.9$&$\{-0.070,0.305,-0.337,0.130,-0.562,0.675\}$\\
        &$2^{+}$
        &$11453.9$&$\{0.926,0.378\}$&$\bar{D}^{*}{\Upsilon}$~($94.9\%$)\\
        &
        &$11747.6$&$\{-0.378,0.926\}$\\
        \midrule[1pt]
        $sb\bar{c}\bar{b}$&$0^{+}$
        &$11355.0$&$\{0.816,0.366,0.364,0.262\}$&$\bar{D}_{s}{\eta}_{b}$~($96.7\%$)\\
        &
        &$11539.4$&$\{0.468,-0.840,-0.008,-0.275\}$&$\bar{D}_{s}^{*}{\Upsilon}$~($84.7\%$)\\
        &
        &$11760.7$&$\{0.159,-0.124,-0.714,0.670\}$\\
        &
        &$11836.7$&$\{0.300,0.381,-0.598,-0.638\}$\\
        &$1^{+}$
        &$11416.4$&$\{0.698,-0.405,0.407,0.296,-0.219,0.218\}$&$\bar{D}_{s}{\Upsilon}$~($96.4\%$)\\
        &
        &$11496.1$&$\{-0.615,-0.437,0.522,-0.216,-0.238,0.235\}$&$\bar{D}_{s}^{*}{\eta}_{b}$~($94.3\%$)\\
        &
        &$11549.1$&$\{-0.029,-0.698,-0.651,-0.027,-0.229,-0.185\}$&$\bar{D}_{s}^{*}{\Upsilon}$~($89.9\%$)\\
        &
        &$11785.1$&$\{0.071,-0.248,-0.151,-0.206,0.717,0.595\}$\\
        &
        &$11802.8$&$\{-0.354,-0.087,-0.019,0.899,0.239,0.024\}$\\
        &
        &$11833.6$&$\{-0.066,0.296,-0.339,0.115,-0.520,0.713\}$\\
        &$2^{+}$
        &$11558.4$&$\{0.926,0.378\}$&$\bar{D}_{s}^{*}{\Upsilon}$~($94.9\%$)\\
        &
        &$11832.7$&$\{-0.378,0.926\}$\\
        \bottomrule[1pt]
        \bottomrule[1pt]
    \end{tabular}
\end{table*}
%

\begin{table*}
    \centering
    \caption{The eigenvectors of the $nc\bar{c}\bar{b}$ tetraquarks. The masses are all in units of MeV.}
    \label{table:eigenvector:nccb}
    \begin{tabular}{ccccccccccc}
        \toprule[1pt]
        \toprule[1pt]
        \multirow{2}{*}{System}&\multirow{2}{*}{$J^{P}$}&\multirow{2}{*}{Mass}&\multicolumn{4}{c}{$n\bar{c}{\otimes}c\bar{b}$}&\multicolumn{4}{c}{$n\bar{b}{\otimes}c\bar{c}$}\\
        \cmidrule(lr){4-7}
        \cmidrule(lr){8-11}
        &&
        &$\bar{D}^{*}{B}_{c}^{*}$&$\bar{D}^{*}{B}_{c}$&$\bar{D}{B}_{c}^{*}$&$\bar{D}{B}_{c}$
        &${B}^{*}{J/\psi}$&${B}^{*}\eta_{c}$&${B}{J/\psi}$&${B}\eta_{c}$\\
        \midrule[1pt]
        $nc\bar{c}\bar{b}$&$0^{+}$
        &$8123.2$&$-0.086$&&&$0.947$
        &$\-0.277$&&&$0.473$\\
        &
        &$8281.2$&$0.649$&&&$0.286$
        &$-0.254$&&&$-0.801$\\
        &
        &$8388.0$&$0.709$&&&$-0.113$
        &$0.291$&&&$0.314$\\
        &
        &$8466.4$&$-0.263$&&&$0.090$
        &$0.880$&&&$-0.187$\\
        &$1^{+}$
        &$8190.4$&$-0.057$&$0.129$&$0.947$&
        &$0.219$&$0.451$&$0.252$\\
        &
        &$8268.0$&$0.108$&$-0.915$&$0.224$&
        &$0.337$&$-0.415$&$-0.192$\\
        &
        &$8312.6$&$0.770$&$0.233$&$0.123$&
        &$-0.048$&$-0.576$&$0.547$\\
        &
        &$8398.9$&$0.610$&$-0.165$&$-0.116$&
        &$-0.143$&$0.476$&$-0.137$\\
        &
        &$8435.7$&$0.089$&$0.252$&$0.146$&
        &$-0.015$&$-0.245$&$-0.762$\\
        &
        &$8457.3$&$-0.113$&$-0.039$&$0.061$&
        &$-0.903$&$-0.086$&$-0.006$\\
        &$2^{+}$
        &$8335.1$&$0.956$&&&
        &$0.595$\\
        &
        &$8468.9$&$0.293$&&&
        &$-0.804$\\
        \bottomrule[1pt]
        \bottomrule[1pt]
    \end{tabular}
\end{table*}
%
\begin{table*}
    \centering
    \caption{The eigenvectors of the $sc\bar{c}\bar{b}$ tetraquarks. The masses are all in units of MeV.}
    \label{table:eigenvector:sccb}
    \begin{tabular}{ccccccccccc}
        \toprule[1pt]
        \toprule[1pt]
        \multirow{2}{*}{System}&\multirow{2}{*}{$J^{P}$}&\multirow{2}{*}{Mass}&\multicolumn{4}{c}{$s\bar{c}{\otimes}c\bar{b}$}&\multicolumn{4}{c}{$s\bar{b}{\otimes}c\bar{c}$}\\
        \cmidrule(lr){4-7}
        \cmidrule(lr){8-11}
        &&
        &$\bar{D}_{s}^{*}{B}_{c}^{*}$&$\bar{D}_{s}^{*}{B}_{c}$&$\bar{D}_{s}{B}_{c}^{*}$&$\bar{D}_{s}{B}_{c}$
        &${B}_{s}^{*}{J/\psi}$&${B}_{s}^{*}\eta_{c}$&${B}_{s}{J/\psi}$&${B}_{s}\eta_{c}$\\
        \midrule[1pt]
        $sc\bar{c}\bar{b}$&$0^{+}$
        &$8220.1$&$-0.100$&&&$0.934$
        &$-0.276$&&&$0.508$\\
        &
        &$8375.2$&$-0.612$&&&$-0.326$
        &$0.279$&&&$0.796$\\
        &
        &$8488.1$&$-0.720$&&&$0.100$
        &$-0.337$&&&$-0.261$\\
        &
        &$8561.0$&$-0.311$&&&$0.109$
        &$0.856$&&&$-0.199$\\
        &$1^{+}$
        &$8287.8$&$-0.064$&$0.145$&$0.933$&
        &$0.217$&$0.483$&$0.262$\\
        &
        &$8368.4$&$0.106$&$-0.897$&$0.257$&
        &$0.361$&$-0.419$&$-0.200$\\
        &
        &$8407.5$&$-0.726$&$-0.238$&$-0.135$&
        &$0.049$&$0.581$&$-0.595$\\
        &
        &$8498.8$&$0.649$&$-0.163$&$-0.107$&
        &$-0.201$&$0.416$&$-0.133$\\
        &
        &$8531.2$&$-0.092$&$-0.294$&$-0.171$&
        &$0.065$&$0.268$&$0.721$\\
        &
        &$8551.1$&$-0.168$&$-0.062$&$0.072$&
        &$-0.880$&$-0.096$&$0.033$\\
        &$2^{+}$
        &$8434.2$&$0.938$&&&
        &$0.639$\\
        &
        &$8566.3$&$0.347$&&&
        &$-0.769$\\
        \bottomrule[1pt]
        \bottomrule[1pt]
    \end{tabular}
\end{table*}
%
\begin{table*}
    \centering
    \caption{The eigenvectors of the $nb\bar{c}\bar{b}$ tetraquarks. The masses are all in units of MeV.}
    \label{table:eigenvector:nbcb}
    \begin{tabular}{ccccccccccc}
        \toprule[1pt]
        \toprule[1pt]
        \multirow{2}{*}{System}&\multirow{2}{*}{$J^{P}$}&\multirow{2}{*}{Mass}&\multicolumn{4}{c}{$n\bar{c}{\otimes}b\bar{b}$}&\multicolumn{4}{c}{$n\bar{b}{\otimes}b\bar{c}$}\\
        \cmidrule(lr){4-7}
        \cmidrule(lr){8-11}
        &&
        &$\bar{D}^{*}{\Upsilon}$&$\bar{D}^{*}{\eta}_{b}$&$\bar{D}{\Upsilon}$&$\bar{D}{\eta}_{b}$
        &${B}^{*}\bar{B}_{c}^{*}$&${B}^{*}\bar{B}_{c}$&${B}\bar{B}_{c}^{*}$&${B}\bar{B}_{c}$\\
        \midrule[1pt]
        $nb\bar{c}\bar{b}$&$0^{+}$
        &$11253.4$&$-0.050$&&&$0.983$
        &$-0.356$&&&$0.322$\\
        &
        &$11438.6$&$-0.931$&&&$-0.094$
        &$0.257$&&&$0.577$\\
        &
        &$11673.1$&$0.361$&&&$-0.099$
        &$0.197$&&&$0.723$\\
        &
        &$11748.7$&$0.019$&&&$-0.119$
        &$-0.877$&&&$0.201$\\
        &$1^{+}$
        &$11314.8$&$-0.005$&$0.072$&$0.981$&
        &$0.288$&$0.289$&$0.278$\\
        &
        &$11392.4$&$0.037$&$-0.972$&$0.091$&
        &$0.361$&$-0.289$&$-0.236$\\
        &
        &$11446.6$&$-0.954$&$-0.047$&$-0.005$&
        &$0.008$&$0.420$&$-0.425$\\
        &
        &$11697.5$&$0.292$&$-0.056$&$-0.044$&
        &$0.047$&$0.689$&$-0.403$\\
        &
        &$11719.1$&$-0.054$&$-0.192$&$-0.126$&
        &$-0.133$&$0.424$&$0.710$\\
        &
        &$11744.9$&$0.028$&$-0.083$&$0.108$&
        &$-0.876$&$-0.048$&$-0.139$\\
        &$2^{+}$
        &$11453.9$&$0.974$&&&
        &$0.538$\\
        &
        &$11747.6$&$0.226$&&&
        &$-0.843$\\
        \bottomrule[1pt]
        \bottomrule[1pt]
    \end{tabular}
\end{table*}
%
\begin{table*}
    \centering
    \caption{The eigenvectors of the $sb\bar{c}\bar{b}$ tetraquarks. The masses are all in units of MeV.}
    \label{table:eigenvector:sbcb}
    \begin{tabular}{ccccccccccc}
        \toprule[1pt]
        \toprule[1pt]
        \multirow{2}{*}{System}&\multirow{2}{*}{$J^{P}$}&\multirow{2}{*}{Mass}&\multicolumn{4}{c}{$s\bar{c}{\otimes}b\bar{b}$}&\multicolumn{4}{c}{$s\bar{b}{\otimes}b\bar{c}$}\\
        \cmidrule(lr){4-7}
        \cmidrule(lr){8-11}
        &&
        &$\bar{D}_{s}^{*}{\Upsilon}$&$\bar{D}_{s}^{*}{\eta}_{b}$&$\bar{D}_{s}{\Upsilon}$&$\bar{D}_{s}{\eta}_{b}$
        &${B}_{s}^{*}\bar{B}_{c}^{*}$&${B}_{s}^{*}\bar{B}_{c}$&${B}_{s}\bar{B}_{c}^{*}$&${B}_{s}\bar{B}_{c}$\\
        \midrule[1pt]
        $sb\bar{c}\bar{b}$&$0^{+}$
        &$11355.0$&$-0.048$&&&$0.984$
        &$-0.356$&&&$0.321$\\
        &
        &$11539.4$&$-0.920$&&&$-0.095$
        &$0.263$&&&$0.598$\\
        &
        &$11760.7$&$0.389$&&&$-0.102$
        &$0.152$&&&$0.714$\\
        &
        &$11836.7$&$0.001$&&&$-0.115$
        &$-0.884$&&&$0.171$\\
        &$1^{+}$
        &$11416.4$&$0.001$&$0.066$&$0.982$&
        &$0.290$&$0.281$&$0.283$\\
        &
        &$11496.1$&$0.048$&$-0.971$&$0.085$&
        &$0.361$&$-0.302$&$-0.231$\\
        &
        &$11549.1$&$-0.948$&$-0.060$&$0.005$&
        &$0.009$&$0.426$&$-0.437$&$$\\
        &
        &$11785.1$&$0.306$&$-0.048$&$-0.039$&
        &$0.106$&$0.666$&$-0.417$\\
        &
        &$11802.8$&$0.046$&$0.197$&$0.129$&
        &$0.127$&$-0.452$&$-0.691$\\
        &
        &$11833.6$&$0.054$&$-0.088$&$0.106$&
        &$-0.871$&$-0.012$&$-0.158$\\
        &$2^{+}$
        &$11558.4$&$0.974$&&&
        &$0.538$\\
        &
        &$11832.7$&$0.226$&&&
        &$-0.843$\\
        \bottomrule[1pt]
        \bottomrule[1pt]
    \end{tabular}
\end{table*}
%

\begin{table*}
    \centering
    \caption{The values of $k\cdot|c_{i}|^2$ for the $nc\bar{c}\bar{b}$ tetraquarks (in unit of MeV).}
    \label{table:kc_i^2:nccb}
    \begin{tabular}{ccccccccccccc}
        \toprule[1pt]
        \toprule[1pt]
        \multirow{2}{*}{System}&\multirow{2}{*}{$J^{P}$}&\multirow{2}{*}{Mass}&\multicolumn{4}{c}{$n\bar{c}{\otimes}c\bar{b}$}&\multicolumn{4}{c}{$n\bar{b}{\otimes}c\bar{c}$}\\
        \cmidrule(lr){4-7}
        \cmidrule(lr){8-11}
        &&
        &$\bar{D}^{*}{B}_{c}^{*}$&$\bar{D}^{*}{B}_{c}$&$\bar{D}{B}_{c}^{*}$&$\bar{D}{B}_{c}$
        &${B}^{*}{J/\psi}$&${B}^{*}\eta_{c}$&${B}{J/\psi}$&${B}\eta_{c}$\\
        \midrule[1pt]
        $nc\bar{c}\bar{b}$&$0^{+}$
        &$8123.2$&$\times$&&&$\times$
        &$\times$&&&$\times$\\
        &
        &$8281.2$&$\times$&&&$52.3$
        &$\times$&&&$170.0$\\
        &
        &$8388.0$&$179.3$&&&$10.9$
        &$\times$&&&$68.6$\\
        &
        &$8466.4$&$42.3$&&&$8.0$
        &$325.1$&&&$31.0$\\
        &$1^{+}$
        &$8190.4$&$\times$&$\times$&$\times$&
        &$\times$&$\times$&$\times$\\
        &
        &$8268.0$&$\times$&$\times$&$21.5$&
        &$\times$&$\times$&$\times$\\
        &
        &$8312.6$&$\times$&$16.1$&$8.5$&
        &$\times$&$43.5$&$\times$\\
        &
        &$8398.9$&$149.1$&$16.2$&$10.3$&
        &$\times$&$133.8$&$5.6$\\
        &
        &$8435.7$&$4.1$&$43.7$&$17.8$&
        &$0.1$&$42.1$&$280.3$\\
        &
        &$8457.3$&$7.5$&$1.1$&$3.2$&
        &$305.7$&$5.6$&$0.02$\\
        &$2^{+}$
        &$8335.1$&$\times$&&&
        &$\times$\\
        &
        &$8468.9$&$52.8$&&&
        &$278.7$\\
        \bottomrule[1pt]
        \bottomrule[1pt]
    \end{tabular}
\end{table*}
%
\begin{table*}
    \centering
    \caption{The values of $k\cdot|c_{i}|^2$ for the $sc\bar{c}\bar{b}$ tetraquarks in the $s\bar{c}{\otimes}c\bar{b}$ configuration (in unit of MeV).}
    \label{table:kc_i^2:sccb}
    \begin{tabular}{ccccccccccccc}
        \toprule[1pt]
        \toprule[1pt]
        \multirow{2}{*}{System}&\multirow{2}{*}{$J^{P}$}&\multirow{2}{*}{Mass}&\multicolumn{4}{c}{$s\bar{c}{\otimes}c\bar{b}$}&\multicolumn{4}{c}{$s\bar{b}{\otimes}c\bar{c}$}\\
        \cmidrule(lr){4-7}
        \cmidrule(lr){8-11}
        &&
        &$\bar{D}_{s}^{*}{B}_{c}^{*}$&$\bar{D}_{s}^{*}{B}_{c}$&$\bar{D}_{s}{B}_{c}^{*}$&$\bar{D}_{s}{B}_{c}$
        &${B}_{s}^{*}{J/\psi}$&${B}_{s}^{*}\eta_{c}$&${B}_{s}{J/\psi}$&${B}_{s}\eta_{c}$\\
        \midrule[1pt]
        $sc\bar{c}\bar{b}$&$0^{+}$
        &$8220.1$&$\times$&&&$\times$
        &$\times$&&&$\times$\\
        &
        &$8375.2$&$\times$&&&$67.5$
        &$\times$&&&$196.4$\\
        &
        &$8488.1$&$180.3$&&&$8.8$
        &$\times$&&&$49.8$\\
        &
        &$8561.0$&$57.6$&&&$11.9$
        &$321.4$&&&$35.9$\\
        &$1^{+}$
        &$8287.8$&$\times$&$\times$&$\times$&
        &$\times$&$\times$&$\times$\\
        &
        &$8368.4$&$\times$&$\times$&$28.7$&
        &$\times$&$\times$&$\times$\\
        &
        &$8407.5$&$\times$&$14.4$&$10.2$&
        &$\times$&$61.9$&$\times$\\
        &
        &$8498.8$&$165.9$&$16.0$&$8.8$&
        &$\times$&$107.8$&$6.5$\\
        &
        &$8531.2$&$4.3$&$58.9$&$24.5$&
        &$1.2$&$51.7$&$268.0$\\
        &
        &$8551.1$&$16.0$&$2.8$&$4.6$&
        &$303.7$&$7.1$&$0.7$\\
        &$2^{+}$
        &$8434.2$&$\times$&&&
        &$\times$\\
        &
        &$8566.3$&$73.5$&&&
        &$273.3$\\
        \bottomrule[1pt]
        \bottomrule[1pt]
    \end{tabular}
\end{table*}
%
\begin{table}
    \centering
    \caption{The partial width ratios for the $nc\bar{c}\bar{b}$ tetraquarks decay into $n\bar{c}{\otimes}c\bar{b}$ modes. For each state, we choose one mode as the reference channel, and the partial width ratios of the other channels are calculated relative to this channel. The masses are all in unit of MeV.}
    \label{table:R:nccb:13x24}
    \begin{tabular}{ccccccccccccc}
        \toprule[1pt]
        \toprule[1pt]
        System&$J^{P}$&Mass&$\bar{D}^{*}{B}_{c}^{*}$&$\bar{D}^{*}{B}_{c}$&$\bar{D}{B}_{c}^{*}$&$\bar{D}{B}_{c}$\\
        \midrule[1pt]
        $nc\bar{c}\bar{b}$&$0^{+}$
        &$8123.2$&$\times$&&&$\times$\\
        &
        &$8281.2$&$\times$&&&$1$\\
        &
        &$8388.0$&$16.4$&&&$1$\\
        &
        &$8466.4$&$5.3$&&&$1$\\
        &$1^{+}$
        &$8190.4$&$\times$&$\times$&$\times$\\
        &
        &$8268.0$&$\times$&$\times$&$1$\\
        &
        &$8312.6$&$\times$&$1.9$&$1$\\
        &
        &$8398.9$&$14.5$&$1.6$&$1$\\
        &
        &$8435.7$&$0.2$&$2.5$&$1$\\
        &
        &$8457.3$&$2.3$&$0.3$&$1$\\
        &$2^{+}$
        &$8335.1$&$\times$\\
        &
        &$8468.9$&$1$\\
        \bottomrule[1pt]
        \bottomrule[1pt]
    \end{tabular}
\end{table}
%
\begin{table}
    \centering
    \caption{The partial width ratios for the $nc\bar{c}\bar{b}$ tetraquarks decay into $n\bar{b}{\otimes}c\bar{c}$ modes. For each state, we choose one mode as the reference channel, and the partial width ratios of the other channels are calculated relative to this channel. The masses are all in unit of MeV.}
    \label{table:R:nccb:14x23}
    \begin{tabular}{ccccccccccccc}
        \toprule[1pt]
        \toprule[1pt]
        System&$J^{P}$&Mass&${B}^{*}{J/\psi}$&${B}^{*}\eta_{c}$&${B}{J/\psi}$&${B}\eta_{c}$\\
        \midrule[1pt]
        $nc\bar{c}\bar{b}$&$0^{+}$
        &$8123.2$&$\times$&&&$\times$\\
        &
        &$8281.2$&$\times$&&&$1$\\
        &
        &$8388.0$&$\times$&&&$1$\\
        &
        &$8466.4$&$10.5$&&&$1$\\
        &$1^{+}$
        &$8190.4$&$\times$&$\times$&$\times$\\
        &
        &$8268.0$&$\times$&$\times$&$\times$\\
        &
        &$8312.6$&$\times$&$1$&$\times$\\
        &
        &$8398.9$&$\times$&$1$&$0.04$\\
        &
        &$8435.7$&$0.001$&$1$&$6.7$\\
        &
        &$8457.3$&$54.3$&$1$&$0.003$\\
        &$2^{+}$
        &$8335.1$&$\times$\\
        &
        &$8468.9$&$1$\\
        \bottomrule[1pt]
        \bottomrule[1pt]
    \end{tabular}
\end{table}
%
\begin{table}
    \centering
    \caption{The partial width ratios for the $sc\bar{c}\bar{b}$ tetraquarks decay into $s\bar{c}{\otimes}c\bar{b}$ modes. For each state, we choose one mode as the reference channel, and the partial width ratios of the other channels are calculated relative to this channel. The masses are all in unit of MeV.}
    \label{table:R:sccb:13x24}
    \begin{tabular}{ccccccccccccc}
        \toprule[1pt]
        \toprule[1pt]
        System&$J^{P}$&Mass&$\bar{D}_{s}^{*}{B}_{c}^{*}$&$\bar{D}_{s}^{*}{B}_{c}$&$\bar{D}_{s}{B}_{c}^{*}$&$\bar{D}_{s}{B}_{c}$\\
        \midrule[1pt]
        $sc\bar{c}\bar{b}$&$0^{+}$
        &$8220.1$&$\times$&&&$\times$\\
        &
        &$8375.2$&$\times$&&&$1$\\
        &
        &$8488.1$&$20.5$&&&$1$\\
        &
        &$8561.0$&$4.8$&&&$1$\\
        &$1^{+}$
        &$8287.8$&$\times$&$\times$&$\times$\\
        &
        &$8368.4$&$\times$&$\times$&$1$\\
        &
        &$8407.5$&$\times$&$1.4$&$1$\\
        &
        &$8498.8$&$18.9$&$1.8$&$1$\\
        &
        &$8531.2$&$0.2$&$2.4$&$1$\\
        &
        &$8551.1$&$3.5$&$0.6$&$1$\\
        &$2^{+}$
        &$8434.2$&$\times$\\
        &
        &$8566.3$&$1$\\
        \bottomrule[1pt]
        \bottomrule[1pt]
    \end{tabular}
\end{table}
%
\begin{table}
    \centering
    \caption{The partial width ratios for the $sc\bar{c}\bar{b}$ tetraquarks decay into $s\bar{b}{\otimes}c\bar{c}$ modes. For each state, we choose one mode as the reference channel, and the partial width ratios of the other channels are calculated relative to this channel. The masses are all in unit of MeV.}
    \label{table:R:sccb:14x23}
    \begin{tabular}{ccccccccccccc}
        \toprule[1pt]
        \toprule[1pt]
        System&$J^{P}$&Mass&${B}_{s}^{*}{J/\psi}$&${B}_{s}^{*}\eta_{c}$&${B}_{s}{J/\psi}$&${B}_{s}\eta_{c}$\\
        \midrule[1pt]
        $sc\bar{c}\bar{b}$&$0^{+}$
        &$8220.1$&$\times$&&&$\times$\\
        &
        &$8375.2$&$\times$&&&$1$\\
        &
        &$8488.1$&$\times$&&&$1$\\
        &
        &$8561.0$&$8.9$&&&$1$\\
        &$1^{+}$
        &$8287.8$&$\times$&$\times$&$\times$\\
        &
        &$8368.4$&$\times$&$\times$&$\times$\\
        &
        &$8407.5$&$\times$&$1$&$\times$\\
        &
        &$8498.8$&$\times$&$1$&$0.1$\\
        &
        &$8531.2$&$0.02$&$1$&$5.2$\\
        &
        &$8551.1$&$42.9$&$1$&$0.1$\\
        &$2^{+}$
        &$8434.2$&$\times$\\
        &
        &$8566.3$&$1$\\
        \bottomrule[1pt]
        \bottomrule[1pt]
    \end{tabular}
\end{table}
%

\begin{table*}
    \centering
    \caption{The values of $k\cdot|c_{i}|^2$ for the $nb\bar{c}\bar{b}$ tetraquarks (in unit of MeV).}
    \label{table:kc_i^2:nbcb}
    \begin{tabular}{ccccccccccccc}
        \toprule[1pt]
        \toprule[1pt]
        \multirow{2}{*}{System}&\multirow{2}{*}{$J^{P}$}&\multirow{2}{*}{Mass}&\multicolumn{4}{c}{$n\bar{c}{\otimes}b\bar{b}$}&\multicolumn{4}{c}{$n\bar{b}{\otimes}b\bar{c}$}\\
        \cmidrule(lr){4-7}
        \cmidrule(lr){8-11}
        &&
        &$\bar{D}^{*}{\Upsilon}$&$\bar{D}^{*}{\eta}_{b}$&$\bar{D}{\Upsilon}$&$\bar{D}{\eta}_{b}$
        &${B}^{*}\bar{B}_{c}^{*}$&${B}^{*}\bar{B}_{c}$&${B}\bar{B}_{c}^{*}$&${B}\bar{B}_{c}$\\
        \midrule[1pt]
        $nb\bar{c}\bar{b}$&$0^{+}$
        &$11253.4$&$\times$&&&$\times$
        &$\times$&&&$\times$\\
        &
        &$11438.6$&$\times$&&&$6.5$
        &$\times$&&&$\times$\\
        &
        &$11673.1$&$109.1$&&&$11.5$
        &$9.5$&&&$432.4$\\
        &
        &$11748.7$&$0.3$&&&$18.2$
        &$543.5$&&&$42.7$\\
        &$1^{+}$
        &$11314.8$&$\times$&$\times$&$\times$&
        &$\times$&$\times$&$\times$\\
        &
        &$11392.4$&$\times$&$\times$&$3.8$&
        &$\times$&$\times$&$\times$\\
        &
        &$11446.6$&$\times$&$0.8$&$0.01$&
        &$\times$&$\times$&$\times$\\
        &
        &$11697.5$&$75.4$&$3.1$&$2.1$&
        &$1.0$&$357.0$&$110.7$\\
        &
        &$11719.1$&$2.7$&$38.6$&$18.0$&
        &$10.1$&$149.3$&$386.9$\\
        &
        &$11744.9$&$0.8$&$7.4$&$13.7$&
        &$529.9$&$2.1$&$16.6$\\
        &$2^{+}$
        &$11453.9$&$\times$&&&
        &$\times$\\
        &
        &$11747.6$&$50.2$&&&
        &$499.3$\\
        \bottomrule[1pt]
        \bottomrule[1pt]
    \end{tabular}
\end{table*}
%
\begin{table*}
    \centering
    \caption{The values of $k\cdot|c_{i}|^2$ for the $sb\bar{c}\bar{b}$ tetraquarks (in unit of MeV).}
    \label{table:kc_i^2:sbcb}
    \begin{tabular}{ccccccccccccc}
        \toprule[1pt]
        \toprule[1pt]
        \multirow{2}{*}{System}&\multirow{2}{*}{$J^{P}$}&\multirow{2}{*}{Mass}&\multicolumn{4}{c}{$s\bar{c}{\otimes}b\bar{b}$}&\multicolumn{4}{c}{$s\bar{b}{\otimes}b\bar{c}$}\\
        \cmidrule(lr){4-7}
        \cmidrule(lr){8-11}
        &&
        &$\bar{D}_{s}^{*}{\Upsilon}$&$\bar{D}_{s}^{*}{\eta}_{b}$&$\bar{D}_{s}{\Upsilon}$&$\bar{D}_{s}{\eta}_{b}$
        &${B}_{s}^{*}\bar{B}_{c}^{*}$&${B}_{s}^{*}\bar{B}_{c}$&${B}_{s}\bar{B}_{c}^{*}$&${B}_{s}\bar{B}_{c}$\\
        \midrule[1pt]
        $sb\bar{c}\bar{b}$&$0^{+}$
        &$11355.0$&$\times$&&&$\times$
        &$\times$&&&$\times$\\
        &
        &$11539.4$&$\times$&&&$6.8$
        &$\times$&&&$\times$\\
        &
        &$11760.7$&$123.5$&&&$12.1$
        &$4.8$&&&$423.6$\\
        &
        &$11836.7$&$0.001$&&&$17.0$
        &$545.9$&&&$31.3$\\
        &$1^{+}$
        &$11416.4$&$\times$&$\times$&$\times$&
        &$\times$&$\times$&$\times$\\
        &
        &$11496.1$&$\times$&$\times$&$3.4$&
        &$\times$&$\times$&$\times$\\
        &
        &$11549.1$&$\times$&$1.3$&$0.01$&
        &$\times$&$\times$&$\times$\\
        &
        &$11785.1$&$81.3$&$2.2$&$1.7$&
        &$4.9$&$330.0$&$118.7$\\
        &
        &$11802.8$&$1.9$&$39.9$&$18.8$&
        &$8.7$&$165.6$&$360.9$\\
        &
        &$11833.6$&$2.8$&$8.3$&$13.2$&
        &$519.6$&$0.1$&$21.7$\\
        &$2^{+}$
        &$11558.4$&$\times$&&&
        &$\times$\\
        &
        &$11832.7$&$49.4$&&&
        &$484.7$\\
        \bottomrule[1pt]
        \bottomrule[1pt]
    \end{tabular}
\end{table*}
%
\begin{table}
    \centering
    \caption{The partial width ratios for the $nb\bar{c}\bar{b}$ tetraquarks decay into $n\bar{c}{\otimes}b\bar{b}$ modes. For each state, we choose one mode as the reference channel, and the partial width ratios of the other channels are calculated relative to this channel. The masses are all in unit of MeV.}
    \label{table:R:nbcb:13x24}
    \begin{tabular}{ccccccccccccc}
        \toprule[1pt]
        \toprule[1pt]
        System&$J^{P}$&Mass&$\bar{D}^{*}{\Upsilon}$&$\bar{D}^{*}{\eta}_{b}$&$\bar{D}{\Upsilon}$&$\bar{D}{\eta}_{b}$\\
        \midrule[1pt]
        $nb\bar{c}\bar{b}$&$0^{+}$
        &$11253.4$&$\times$&&&$\times$\\
        &
        &$11438.6$&$\times$&&&$1$\\
        &
        &$11673.1$&$9.5$&&&$1$\\
        &
        &$11748.7$&$0.02$&&&$1$\\
        &$1^{+}$
        &$11314.8$&$\times$&$\times$&$\times$\\
        &
        &$11392.4$&$\times$&$\times$&$1$\\
        &
        &$11446.6$&$\times$&$59.0$&$1$\\
        &
        &$11697.5$&$35.6$&$1.5$&$1$\\
        &
        &$11719.1$&$0.1$&$2.1$&$1$\\
        &
        &$11744.9$&$0.1$&$0.5$&$1$\\
        &$2^{+}$
        &$11453.9$&$\times$\\
        &
        &$11747.6$&$1$\\
        \bottomrule[1pt]
        \bottomrule[1pt]
    \end{tabular}
\end{table}
%
\begin{table}
    \centering
    \caption{The partial width ratios for the $nb\bar{c}\bar{b}$ tetraquarks decay into $n\bar{b}{\otimes}b\bar{c}$ modes. For each state, we choose one mode as the reference channel, and the partial width ratios of the other channels are calculated relative to this channel. The masses are all in unit of MeV.}
    \label{table:R:nbcb:14x23}
    \begin{tabular}{ccccccccccccc}
        \toprule[1pt]
        \toprule[1pt]
        System&$J^{P}$&Mass&${B}^{*}\bar{B}_{c}^{*}$&${B}^{*}\bar{B}_{c}$&${B}\bar{B}_{c}^{*}$&${B}\bar{B}_{c}$\\
        \midrule[1pt]
        $nb\bar{c}\bar{b}$&$0^{+}$
        &$11253.4$&$\times$&&&$\times$\\
        &
        &$11438.6$&$\times$&&&$\times$\\
        &
        &$11673.1$&$0.02$&&&$1$\\
        &
        &$11748.7$&$12.7$&&&$1$\\
        &$1^{+}$
        &$11314.8$&$\times$&$\times$&$\times$\\
        &
        &$11392.4$&$\times$&$\times$&$\times$\\
        &
        &$11446.6$&$\times$&$\times$&$\times$\\
        &
        &$11697.5$&$0.01$&$3.2$&$1$\\
        &
        &$11719.1$&$0.03$&$0.4$&$1$\\
        &
        &$11744.9$&$31.9$&$0.1$&$1$\\
        &$2^{+}$
        &$11453.9$&$\times$\\
        &
        &$11747.6$&$1$\\
        \bottomrule[1pt]
        \bottomrule[1pt]
    \end{tabular}
\end{table}
%
\begin{table}
    \centering
    \caption{The partial width ratios for the $sb\bar{c}\bar{b}$ tetraquarks decay into $s\bar{c}{\otimes}b\bar{b}$ modes. For each state, we choose one mode as the reference channel, and the partial width ratios of the other channels are calculated relative to this channel. The masses are all in unit of MeV.}
    \label{table:R:sbcb:13x24}
    \begin{tabular}{ccccccccccccc}
        \toprule[1pt]
        \toprule[1pt]
        System&$J^{P}$&Mass&$\bar{D}_{s}^{*}{\Upsilon}$&$\bar{D}_{s}^{*}{\eta}_{b}$&$\bar{D}_{s}{\Upsilon}$&$\bar{D}_{s}{\eta}_{b}$\\
        \midrule[1pt]
        $sb\bar{c}\bar{b}$&$0^{+}$
        &$11355.0$&$\times$&&&$\times$\\
        &
        &$11539.4$&$\times$&&&$1$\\
        &
        &$11760.7$&$10.2$&&&$1$\\
        &
        &$11836.7$&$0.00003$&&&$1$\\
        &$1^{+}$
        &$11416.4$&$\times$&$\times$&$\times$\\
        &
        &$11496.1$&$\times$&$\times$&$1$\\
        &
        &$11549.1$&$\times$&$99.7$&$1$\\
        &
        &$11785.1$&$47.8$&$1.3$&$1$\\
        &
        &$11802.8$&$0.1$&$2.1$&$1$\\
        &
        &$11833.6$&$0.2$&$0.6$&$1$\\
        &$2^{+}$
        &$11558.4$&$\times$\\
        &
        &$11832.7$&$1$\\
        \bottomrule[1pt]
        \bottomrule[1pt]
    \end{tabular}
\end{table}
%
\begin{table}
    \centering
    \caption{The partial width ratios for the $sb\bar{c}\bar{b}$ tetraquarks decay into $s\bar{b}{\otimes}b\bar{c}$ modes. For each state, we choose one mode as the reference channel, and the partial width ratios of the other channels are calculated relative to this channel. The masses are all in unit of MeV.}
    \label{table:R:sbcb:14x23}
    \begin{tabular}{ccccccccccccc}
        \toprule[1pt]
        \toprule[1pt]
        System&$J^{P}$&Mass&${B}_{s}^{*}\bar{B}_{c}^{*}$&${B}_{s}^{*}\bar{B}_{c}$&${B}_{s}\bar{B}_{c}^{*}$&${B}_{s}\bar{B}_{c}$\\
        \midrule[1pt]
        $sb\bar{c}\bar{b}$&$0^{+}$
        &$11355.0$&$\times$&&&$\times$\\
        &
        &$11539.4$&$\times$&&&$\times$\\
        &
        &$11760.7$&$0.01$&&&$1$\\
        &
        &$11836.7$&$17.4$&&&$1$\\
        &$1^{+}$
        &$11416.4$&$\times$&$\times$&$\times$\\
        &
        &$11496.1$&$\times$&$\times$&$\times$\\
        &
        &$11549.1$&$\times$&$\times$&$\times$\\
        &
        &$11785.1$&$0.04$&$2.8$&$1$\\
        &
        &$11802.8$&$0.02$&$0.5$&$1$\\
        &
        &$11833.6$&$24.0$&$0.01$&$1$\\
        &$2^{+}$
        &$11558.4$&$\times$\\
        &
        &$11832.7$&$1$\\
        \bottomrule[1pt]
        \bottomrule[1pt]
    \end{tabular}
\end{table}

Now we turn to the systems with two different heavy antiquarks.
They are not constrained by Pauli principle.
The mass spectra and eigenvectors are listed in
Tables~\ref{table:mass:nccb+sccb}--\ref{table:mass:nbcb+sbcb}.
We also transform their eigenvectors into the
$q\bar{c}{\otimes}Q\bar{b}$ and $q\bar{b}{\otimes}Q\bar{c}$
configurations, as shown in
Tables~\ref{table:eigenvector:nccb}--\ref{table:eigenvector:sbcb}.
From these tables, we see that some eigenstates couple very strongly
with two $S$-wave mesons.
Taking $nb\bar{c}\bar{b}$ tetraquarks as example, the lowest
eigenstate
\begin{equation}
T(nb\bar{c}\bar{b},11293.4,0^{+}) = 0.98341\bar{D}\eta_{b}
+\cdots\,.
\end{equation}
This state couples almost completely ($96.7\%$) to the
$\bar{D}\eta_{b}$ scattering state.
Thus it may be very broad and hid in the continuum.
Note that this kind of state also exists in the calculation of the
hidden charm tetraquarks and pentaquarks, where the lower mass state
couples strongly to a heavy charmonium and a light
hadron~\cite{Cui:2006mp,Hogaasen:2005jv,Weng:2019ynv}.
Moreover, the $T(nb\bar{c}\bar{b},11314.8,1^{+})$ and
$T(nb\bar{c}\bar{b},11392.4,1^{+})$ states couple very strongly to
$\bar{D}\Upsilon$ and $\bar{D}^{*}\eta_{b}$ channels.
Both the $T(nb\bar{c}\bar{b},11438.6,0^{+})$,
$T(nb\bar{c}\bar{b},11446.6,1^{+})$ and
$T(nb\bar{c}\bar{b},11453.9,2^{+})$ couple very strongly to
$\bar{D}^{*}\Upsilon$.
They may be scattering states.
To draw a definite conclusion, dynamical studies like the complex
scaling methods~\cite{Myo:2020rni,Yang:2020atz} are needed, which is
beyond the present work.
For clarity, we add a fifth column in
Tables~\ref{table:mass:nccb+sccb}--\ref{table:mass:nbcb+sbcb} to
indicate these scattering states.

We plot the relative position of the $qc\bar{c}\bar{b}$ and
$qb\bar{c}\bar{b}$ tetraquarks in
Figs.~\ref{fig:nccb+sccb}--\ref{fig:nbcb+sbcb}.
For comparison, we also plot the possible scattering states.
They are marked with a dagger ($\dagger$), along with the proportion
of their dominant components.
We can easily see that they all lie close to the corresponding
meson-meson thresholds.

After identifying the scattering states, the other states are
genuine tetraquarks.
They all lie far above their $S$-wave decay channel(s).
In particular, the $qb\bar{c}\bar{b}$ tetraquarks lie above all
possible thresholds of two $S$-wave mesons.
Thus they may be broad states.
We also study their decay properties, which can be found in
Tables~\ref{table:kc_i^2:nccb}--\ref{table:R:sbcb:14x23}.
%

\section{Conclusions}
\label{Sec:Conclusion}

Experimentally, lots of $XYZ$ states have been found in this
century.
The singly charmed $X(2900)$ and doubly charmed $T_{cc}^{+}$ were
also observed recently.
For all these system, the exchange of a light meson may be important
since they are composed at least two light (anti)quarks.
It is difficult to determine whether they are loosely bound
molecular states or compact tetraquark states.
The recently discovered $X(6900)$'s are composed of four charm
quarks.
They are very likely compact tetraquarks since their interactions
are mainly provided by gluon exchange.
Similarly, if a triply heavy exotic state is observed in experiment,
it is very likely a compact tetraquark.

In this work, we have systematically studied the triply heavy
tetraquarks in an extended chromomagnetic model, which includes both
colorelectric and chromomagnetic interactions.
Our calculation suggests that the energy level is mainly determined
by the color interaction.
For the $qc\bar{c}\bar{c}$, $qb\bar{c}\bar{c}$ and
$qb\bar{b}\bar{b}$ tetraquarks, the ground states are dominated by
the $6_{c}\otimes\bar{6}_{c}$ component, which is similar to the
fully heavy tetraquarks.
However, the $qc\bar{b}\bar{b}$ tetraquarks are more like the doubly
heavy tetraquarks, where the $6_{c}\otimes\bar{6}_{c}$ component is
mostly in the higher mass states, and the ground states are
dominated by $\bar{3}_{c}\otimes3_{c}$ components.
We find no stable state which lie below the thresholds of two
pseudoscalar mesons, in consistence with the results of Ref.~\cite{Lu:2021kut}.
The lowest axial-vector states with $qQ\bar{b}\bar{b}$ flavor
configuration might be narrow because they lie just above the
thresholds of two pseudoscalar mesons, but they cannot decay into
these channels because of the conservation of the angular momentum
and parity.

With the obtained wave functions, we also calculate the partial
decay rates of the tetraquarks.
We hope that our studies can be of help for the future experimental
searches.
%

\section*{Acknowledgments}

This project was supported by the National Natural Science
Foundation of China (NSFC) under Grants No.~11975033 and
No.~12070131001.
%

\bibliography{myreference}
\end{document}